# Atmospheric Retrieval for Direct Imaging Spectroscopy of Gas Giants In Reflected Light II: Orbital Phase and Planetary Radius

Michael Nayak[1,2,3], Roxana Lupu[1,5], Mark S. Marley[1], Jonathan J. Fortney[4], Tyler Robinson[1,4] and Nikole Lewis[6]

[1]NASA Ames Research Center, Moffett Field, CA 94035; now at Maui High Performance Computing Center (MHPCC). nayak@redskyresearch.org.
[2]Dept. of Earth and Planetary Science, Univ. of California at Santa Cruz, Santa Cruz CA 95064
[3]Red Sky Research, LLC, Edgewood NM 87015
[4]Dept. of Astronomy and Astrophysics, Univ. of California at Santa Cruz, Santa Cruz CA 95064
[5]Bay Area Environmental Research Institute, Moffett Field, CA 94035
[6]Space Telescope Science Institute, Baltimore, MD 21218

## Abstract

Future space-based telescopes, such as the Wide-Field Infrared Survey Telescope (WFIRST), will observe the reflected-light spectra of directly imaged extrasolar planets. Interpretation of such data presents a number of novel challenges, including accounting for unknown planet radius and uncertain stellar illumination phase angle. Here we report on our continued development of Markov Chain Monte Carlo retrieval methods for addressing these issues in the interpretation of such data. Specifically we explore how the unknown planet radius and potentially poorly known observer-planet-star phase angle impacts retrievals of parameters of interest such as atmospheric methane abundance, cloud properties and surface gravity. As expected, the uncertainty in retrieved values is a strong function of signal-to-noise ratio (SNR) of the observed spectra, particularly for low metallicity atmospheres, which lack deep absorption signatures. Meaningful results may only be possible above certain SNR thresholds; for cases across a metallicity range of 1-50 times solar, we find that only an SNR of 20 systematically reproduces close to the correct methane abundance at all phase angles. However, even in cases where the phase angle is poorly known we find that the planet radius can be constrained to within a factor of two. We find that uncertainty in planet radius decreases at phase angles past quadrature, as the highly forward scattering nature of the atmosphere at these geometries limits the possible volume of phase space that relevant parameters can occupy. Finally, we present an estimation of possible improvement that can result from combining retrievals against observations at multiple phase angles.

## 1. Introduction

Transit and radial velocity (RV) surveys have been highly successful in detecting short-period exoplanet systems, and have allowed the compilation of a statistical picture of the bulk properties of inner planetary systems. However, the next frontier in exoplanet studies is space-based direct imaging and spectroscopy using optical-wavelength telescopes, coronagraphs and integral field spectrographs. Such instruments will allow the characterization of colder or self-luminous planets that orbit farther from their parent star. The upcoming Wide-Field Infra-Red Survey Telescope (WFIRST) space telescope will feature a space-based high-contrast coronagraph for imaging and spectroscopic studies of planets around nearby stars (Spergel et al. 2013). It will perform spectroscopy of extrasolar planets in reflected light at spectral resolutions of R~70, in wavelengths ranging from ~600-970 nm. Unlike transit spectroscopy, which is able to probe to atmospheric pressures of ~1 mbar (Kreidberg et al. 2014), or ~1 bar in combination with emission spectra, direct imaging has the potential to probe deeper into the atmosphere, to the pressure at which atmospheric aerosols become optically thick (Morley et al. 2015).

To support the definition of future direct imaging missions and to enhance their science returns, we have been developing a set of tools to characterize gas giant planet atmospheric and physical properties using reflected light spectroscopy, given anticipated instrument parameters from WFIRST. In Lupu et al. (2016), the first in this series (henceforth ***Paper 1***), retrievals of atmospheric methane abundances and basic cloud properties using Markov Chain Monte Carlo (MCMC) techniques were explored, assuming planets with known radii were observed at full phase. ***Paper 1*** largely built on the forward modeling efforts of Marley et al. (1999), and leveraged albedo variations as a function of cloud structure, mass, metallicity, planet phase and star-planet separation by Cahoy et al. (2010). Other contributions in this field have included Sudarsky et al. (2000, 2003) and Burrows et al. (2004). All these studies of reflected light spectra of exoplanets modeled the planets at full phase (***Paper 1***; Marley et al. 1999), an average phase (Sudarsky et al. 2003) or at a set of specified phase angles (Cahoy et al. 2010; Sudarsky et al. 2005) and implicitly assumed that the incident flux and planet size were known.

However during a real observation campaign, several factors that control the reflected flux will be poorly known. First, planets will be observed at a variety of different points along their orbits. Depending on the fidelity with which the orbit is constrained, the star-planet-observer angle (phase angle), and thus the illumination phase, may not be well known. The instantaneous distance of the planet from its star, and thus the incident flux, will almost certainly not be perfectly known. Likewise, planet radii will not be constrained, except by the observed brightness of the planet at a variety of wavelengths and the mass-radius relationship for those planets with masses constrained by radial velocity measurements. Any uncertainties in orbital phase will further obscure planet radius determination.

Figure 1 illustrates the degenerate nature of planet radius with increasing planet phase in scattered light; the brightness of a planet can decrease either with decreasing planet radius or increasing phase. In other words, a large planet at quadrature (phase angle $\alpha$ = 90°) and a smaller planet at full-phase ($\alpha$ = 0°) could not be distinguishable solely by their relative brightness. As we shall show, this degeneracy has a significant effect on the quality of the resulting retrievals on other parameters of interest, including methane abundance and cloud properties.

Our goal in this work is to better explore the bounds of these mutually dependent parameters and determine signal-to-noise (SNR) requirements for scientifically interesting conclusions to be drawn. Simple circular Keplerian orbits are assumed, with constant planet-star separations, in order to initially characterize changes in retrieved molecular abundances, surface gravity and cloud properties with changing phase angles. Inclined orbits require consideration of the degeneracy between inclination and orbital phase, which can result in similar "phase angles" from a reflected light point of view, which we do not undertake here.

The paper is organized as follows: Section 2 provides background on our approach to modeling the phase angle and clouds, our MCMC formulation and our chosen exoplanet test cases of HD 99492c and HD 192310c; Section 3 details MCMC retrieval results and the use of posterior probability plots to extract 68% confidence intervals on parameters of interest. Section 4 contains our discussion of how planet radius, phase angle, methane abundance and cloud properties were constrained in the presence of planet phase and radius uncertainties; finally, Section 5 presents our conclusions.

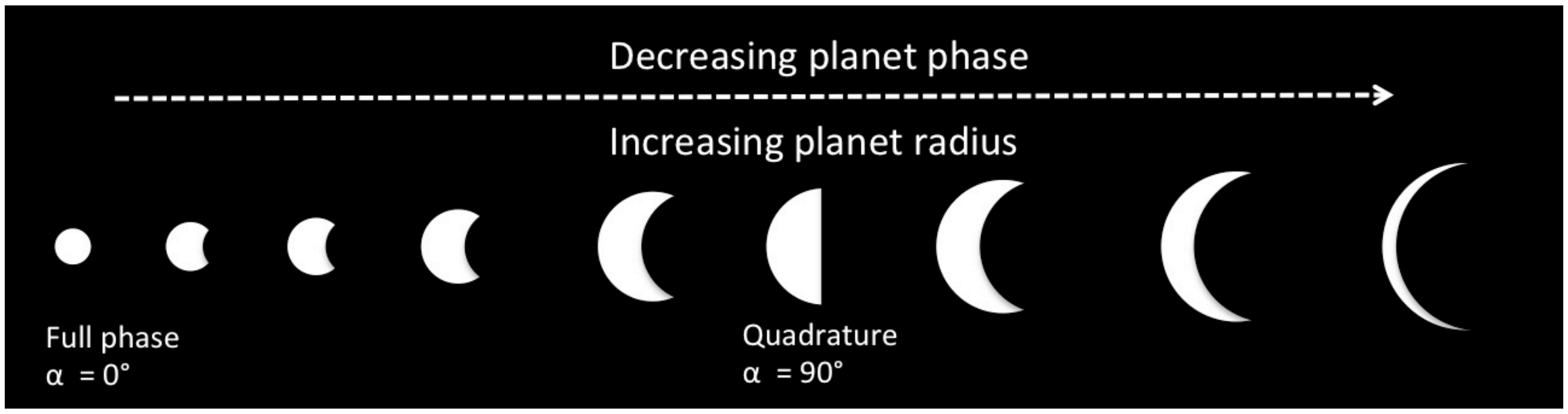

**Figure 1. Illustration of the degenerate relationship between decreasing planet phase (increasing phase angle $\alpha$) and increasing planet radius, in yielding an equivalent scattered flux. If the planet phase is unknown or uncertain, a larger planet at a crescent phase may reflect essentially the same amount of light as a smaller planet at a fuller phase.**

## 2. Background

In this section, we provide a brief overview of some key concepts discussed in the paper. This work builds on previous work by several authors to create models of albedo spectra of extrasolar giant planets; a thorough description may be found in ***Paper 1***. Our initial study reported in that paper represented the first time molecular abundances and cloud properties were simultaneously retrieved using Bayesian inference tools applied to simulated scattered light spectra of cool extrasolar giant planets. Similar applications of Bayesian methods to exoplanet studies include work by Irwin et al. (2008), Madhusudhan & Seager (2009), Madhusudhan et al. (2011, 2014), Benneke & Seager (2012, 2013), Lee et al. (2012, 2013), Line et al. (2013, 2014a), Barstow et al. (2013, 2014, 2015) and to brown dwarfs by Line et al. (2014b, 2015).

### 2.1 Albedo Model

To compute the thermal structure of each model planet's atmosphere, we use a 1D radiative-convective equilibrium model based on that developed for Titan (McKay et al. 1989) and solar system giant planets and exoplanets (Marley et al. 1999; Marley & McKay 1999). The methane opacity at optical wavelengths is taken from Karkoschka (1994) and collision-induced absorption from Freedman et al. (2008). In this paper our test planets are cold and we neglect $H_2O$ opacity. Given a self-consistent model atmospheric profile, we compute an albedo spectrum following the methods described in Cahoy et al. (2010). Cloud scattering is treated with a two-term Heyney-Greenstein function, which captures moderate backscattering and high forward scattering as:

$$p_{TT-HG} = \left(1 - \frac{\bar{g}^2}{4}\right) p_{HG}(\bar{g}, \theta) + \frac{\bar{g}^2}{4} p_{HG}\left(-\frac{\bar{g}}{2}, \theta\right) \quad (1)$$

where:

$$p_{HG}(\bar{g}, \theta) = \frac{1-\bar{g}^2}{(1+\bar{g}^2-2\bar{g}cos\theta)^{1.5}} \quad (2)$$

Here, $\bar{g}$ is the scattering asymmetry factor, which is a retrievable quantity. Integrating over the emergent intensity using Gaussian-Chebyshev quadrature (***Paper 1***, originally from Horak (1950) and Horak & Little (1965)) yields model albedo spectra for the planet. In this paper we treat the phase angle as fully unknown. We allow it to vary from full phase ($\alpha$ = 0°) to $\alpha$ = 135°, at which angle the flux for a Lambertian sphere is <5% of that at full phase. In reality, we will have some constraint on phase angle from radial velocity (RV) observations, which will improve retrievals; our results may therefore be treated as a worst-case scenario (but see Section 4.2). Similarly, even though observations at full phase are not possible for direct imaging, we include that possibility to understand how the quality of retrievals at other phases may relate to those at zero phase, which were elucidated in ***Paper 1***.

### 2.2 MCMC Retrieval Methodology

For our implementation of Markov Chain Monte Carlo (MCMC) methods, we follow the approach developed for massively parallel implementations by Lupu et al. (2016) and use *emcee*, an open-source affine invariant ensemble MCMC sampler (Foreman-Mackey et al. 2013; Goodman & Weare 2010). Given a set of well-chosen bounds on retrievable parameters ("priors"), this approach efficiently samples the parameter space and allows for massively parallel computation. For each retrievable parameter, this implementation employs multiple MCMC chains in parallel. Another multimodel nested sampling algorithm is *Multinest*, which was also utilized in ***Paper 1***; as discussed there, since *Multinest* could favor highly-peaked Gaussian-like distributions, we choose to use *emcee* since it is more agnostic to the shape of the posterior, in case this reveals additional tails or correlations.

As also described in ***Paper 1***, we apply these methods to simulated spectral datasets, computed as in Section 2.1, to retrieve quantities of interest in the interpretation of exoplanet spectra. Table 1 lists all eleven retrievable quantities that are estimated by our Markov Chain Monte Carlo (MCMC) routine. It assumes that the atmosphere's major absorber is solely methane, with $H_2$-He background gas fixed to the solar composition value. Priors are set to allow values to range across six orders of magnitude for methane abundance, 2.5 orders of magnitude for surface gravity and three orders of magnitude for planetary radius (Table 1). Of course, these are extremely large ranges; in reality better constraints are expected. For example, astrometry combined with RV constraints will likely determine the planet mass to within a factor of two. Likewise, the mass-radius relationship trivially demonstrates that a Jupiter mass planet would never have a radius of 100 $R_J$. However, the exploration of a large parameter space can be valuable in permitting a greater number of feasible solutions, enabling a better understanding of the degeneracies inherent in the problem. As an example, all else being equal, lower methane abundance at lower gravity can produce a similar absorption feature to a higher methane abundance at a higher surface gravity (Marley et al. 2014, ***Paper 1***); the use of MCMC to explore this large parameter space may be useful to probabilistically distinguish between the two cases.

**Table 1. Description of retrievable parameters in two-layer cloud model** (Marley et al. 2014)**, as well as ranges of priors used for MCMC runs.**

| Quantity | Description | Priors | Descriptor |
|---|---|---|---|
| $f_{CH4}$ | Molecular abundance of Methane | -8 to -2 | log space |
| g | Surface acceleration due to gravity | 1 - 300 | m/s$^2$ |
| R | Planet Radius | 0.1 - 100 | Jupiter radius |
| $dP_1$ | Pressure difference: Top of lower cloud to Bottom of upper cloud | -2 to 2 | log space |
| $dP_2$ | Pressure difference: Bottom of upper cloud to Top of upper cloud | -2 to 2 | log space |
| $\tau$ | Total optical depth, upper cloud | -2 to 2 | log space |
| $\varpi_1$ | Single scattering albedo, upper cloud | 0.01 to 0.9999 | |
| $\breve{G}$ | Asymmetry Factor | 0.01 to 0.9999 | |
| P | Pressure, top of lower cloud | -2 to 1.5 | log space |
| $\varpi_2$ | Single scattering albedo, lower cloud | 0.01 to 0.9999 | |
| $\Phi$ | Planet phase angle | 0 to 135 | degrees |

Also among the retrievable parameters, exoplanet cloud and haze aerosols are parameterized using an improved version of the simple two-layer cloud model first detailed in Marley et al. (2014). Using the two-layer cloud model, illustrated in Figure 2, we create a model cloud and a noise-free albedo spectrum (Marley et al. 2014). These quantities are also used in ***Paper 1***, which contrasts the two-layer model against simpler one-layer and no cloud models.

To understand how instrumental and astrophysical parameters can affect observed spectra, and include these effects in our retrievals, we apply a noise model to the noise-free spectrum, which includes convolution with an instrument point spread function (PSF) to an appropriate spectral resolution, notionally representative of that expected for WFIRST. Parameters of the noise model are presented in Table 2 and the implementation follows Robinson et al. (2015). Each simulated data point is drawn from a normal distribution, with the mean given by the planet-star flux ratio and the standard distribution given by the noise model. This noisy spectrum then becomes the input to the MCMC retrieval code.

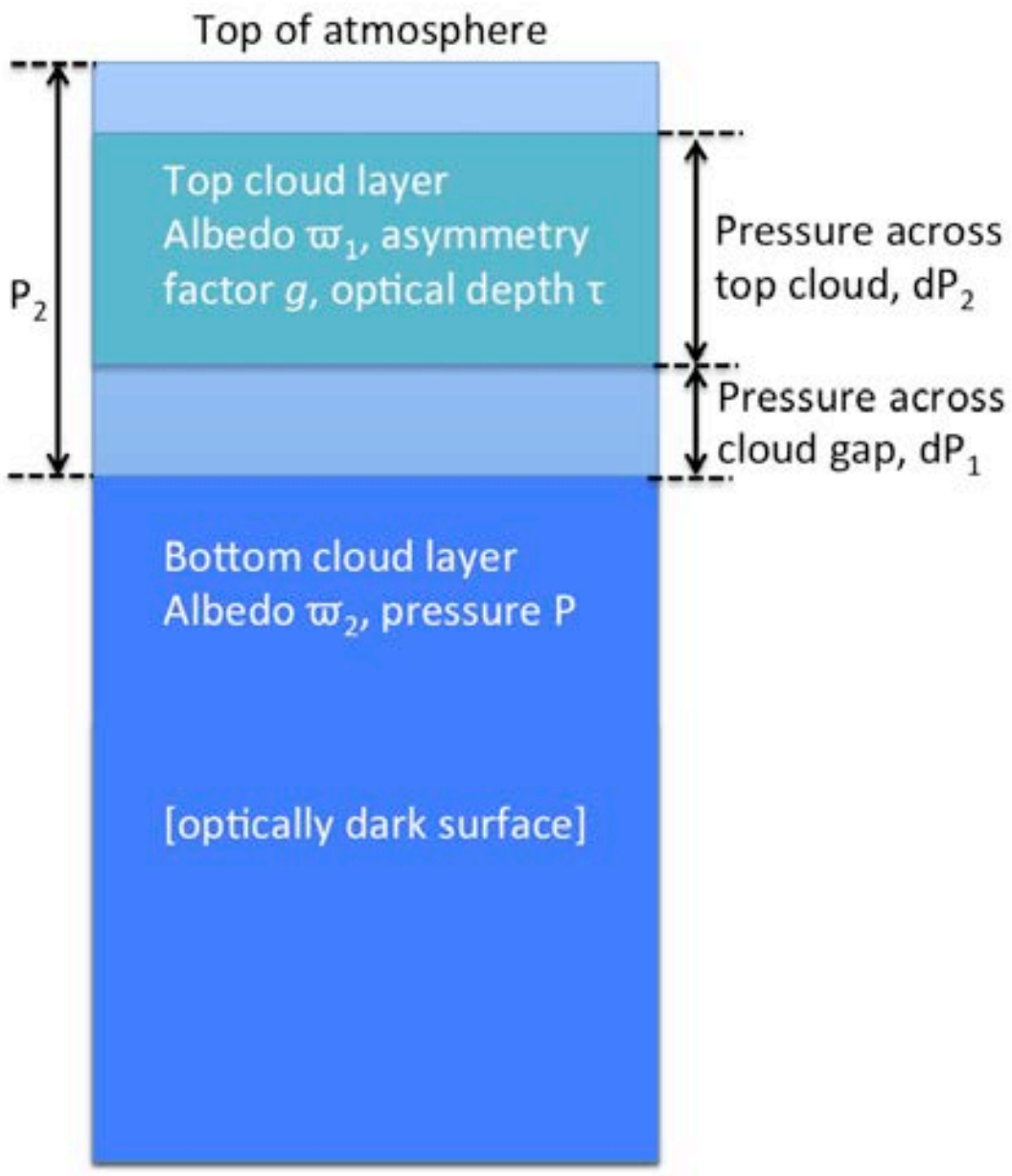

**Figure 2. Illustrative representation of two-layer cloud model employed, after Marley et al. (2014).**

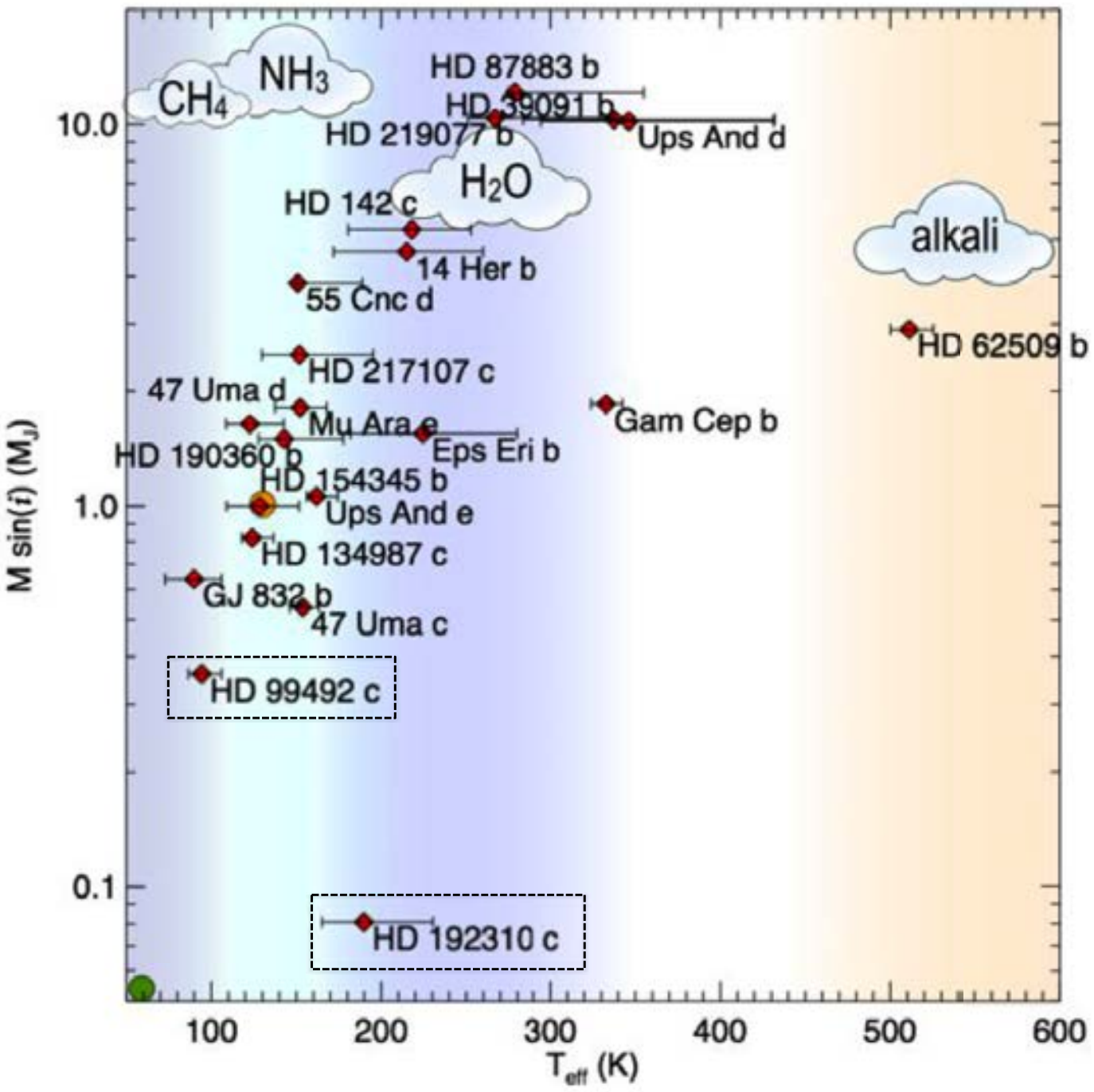

**Figure 3. Candidate target planets favorable for characterization by WFIRST. The *M* sin *i* of each planet, as determined by radial-velocity measurements, is plotted against the planet's estimated effective temperature, accounting for both absorption of incident flux and thermal evolution as described in Marley et al. (2014). Color banding indicates approximate effective temperature regimes where various clouds will dominate the reflected flux signal. Dashed boxes highlight planets discussed in this work. Jupiter is indicated by the gold dot. The green dot indicates the effective temperature (but not the mass, which is lower) of Uranus.**

We make three notes here: firstly, at the time of this writing, the WFIRST coronagraph instrument parameters are still under active study and refinement (Harding et al. 2015). However, our adopted values (Table 2) are meant to be representative of the noise levels that the mission is expected capable of achieving, as it is understood to be in mid-2016. Secondly, the instrument-representative approach employed here differs from ***Paper 1***, which uses a general synthetic noise model. Finally, here we do not attempt to retrieve the atmospheric temperature-pressure profile, as the reflected light spectra are only weakly dependent on the profile. Variations in gravity do alter the scale height and atmospheric density and these effects are accounted for. A future paper in this series will explore atmospheric temperature profile retrievals.

**Table 2. Parameters used in the notional WFIRST noise model. Details on implementation follow** (Robinson et al. 2015).

| Item | WFIRST representative value | Unit |
|---|---|---|
| Dark current | 5.00E-04 | s-1 |
| Telescope diameter | 2.4 | m |
| Read noise | 0.2 | per pixel |
| System throughput | 0.037 | |
| Angular size of lenslet | 0.017 | arcsecond |
| Inner working angle | 2.7 | λ/D |
| Outer working angle | 10 | λ/D |
| Size of photometric aperture | 1.5 | λ/D |
| Contrast floor | 1.00E-10 | |
| Minimum wavelength | 0.6 | µm |
| Maximum wavelength | 0.95 | µm |
| Spectral resolution | R = 70 | |

**Table 3. Truth values for methane abundance, surface gravity, planet radius and cloud pressures for all four test cases in this work.**

| | Planet A | HD 192310c Metallicity 1x | HD 192310c Metallicity 10x | HD 192310c Metallicity 50x |
|---|---|---|---|---|
| log ($fCH_4$) | -3.31 | -3.33 | -2.33 | -1.62 |
| Gravity ($ms^{-2}$) | 7.3 | 3.72 | 3.72 | 3.72 |
| Radius ($R_{Jup}$) | 0.9 | 0.625 | 0.625 | 0.625 |
| log (P2) (top of bottom cloud) | -0.1 | 1.65 | 0.45 | 0.15 |
| log (P3) (top of top cloud) | -0.7 | -2.55 | -2.70 | -2.40 |

## 2.3 Test cases: Synthetic HD 99492c (Planet A) and HD 192310c

Four test cases across two idealized exoplanets encompass our efforts to contrast the relative effects of changing planet mass. Represented in the retrievable parameters by surface gravity and planet radius, the planet mass in turn controls other retrieved properties. Truth-values for all four cases, discussed below, are presented in Table 3.

For a non-solar system test case, ***Paper 1*** used HD 99492c (Marcy et al. 2005); our first test case is also inspired by HD 99492c. The inferred mass of this planet is $0.36 \pm 0.02$ $M_J$ (Meschiari et al. 2010, Table 2); at a semi-major axis of 5.4 AU from its star, models from Fortney et al. (2007) suggest a radius for this planet of ~0.9 $R_J$. Kane et al. (2016) report that HD 99492c is in fact an artifact attributable to variability of the host star and not a planet. Therefore we treat spectra generated for HD 99492c as a synthetic case study to baseline our results against, for a planet almost of Jupiter radius with a methane-dominated atmosphere. To permit observations at all phase angles to be above the WFIRST contrast floor (Table 2), we "relocate" our synthetic planet to 2 AU from its star and 5 parsecs from the telescope; the actual values for HD 99492c are 5.4 AU / 18 parsecs (Marcy et al. 2005). We refer to this synthetic HD 99492c analog as "Planet A" for the remainder of this work.

The second test case explores a planet with an order of magnitude smaller mass (Figure 3). HD 192310c, also known as Gliese 785 c, has $M \sin i$ of 0.076±0.016 $M_J$ (Pepe et al. 2011). Using mass-radius relationships from Fortney et al. (2007) we infer a radius of 0.75 $R_J$. Using the notional parameters in Table 2, this planet will be near the coronagraph inner working angle (IWA) at 600 nm. However, since the WFIRST coronagraphs are still under development, their in-flight performance may yet change. Thus, we choose not to relocate the planet as for HD 99492c / Planet A, and instead use HD 192310c as a test case for situations where a planet lies near enough to the coronagraph IWA to cause the noise to be dominated by stellar leakage. Using this approach, we produce three distinct models of HD 192310c at metallicities of 1x, 10x and 50x times solar abundance, which produce successively deeper methane features (see Figure 6). Figure 4 contrasts model spectra for Planet A and the 50x solar metallicity HD 192310c at $\alpha$ = 0° and $\alpha$ = 90°.

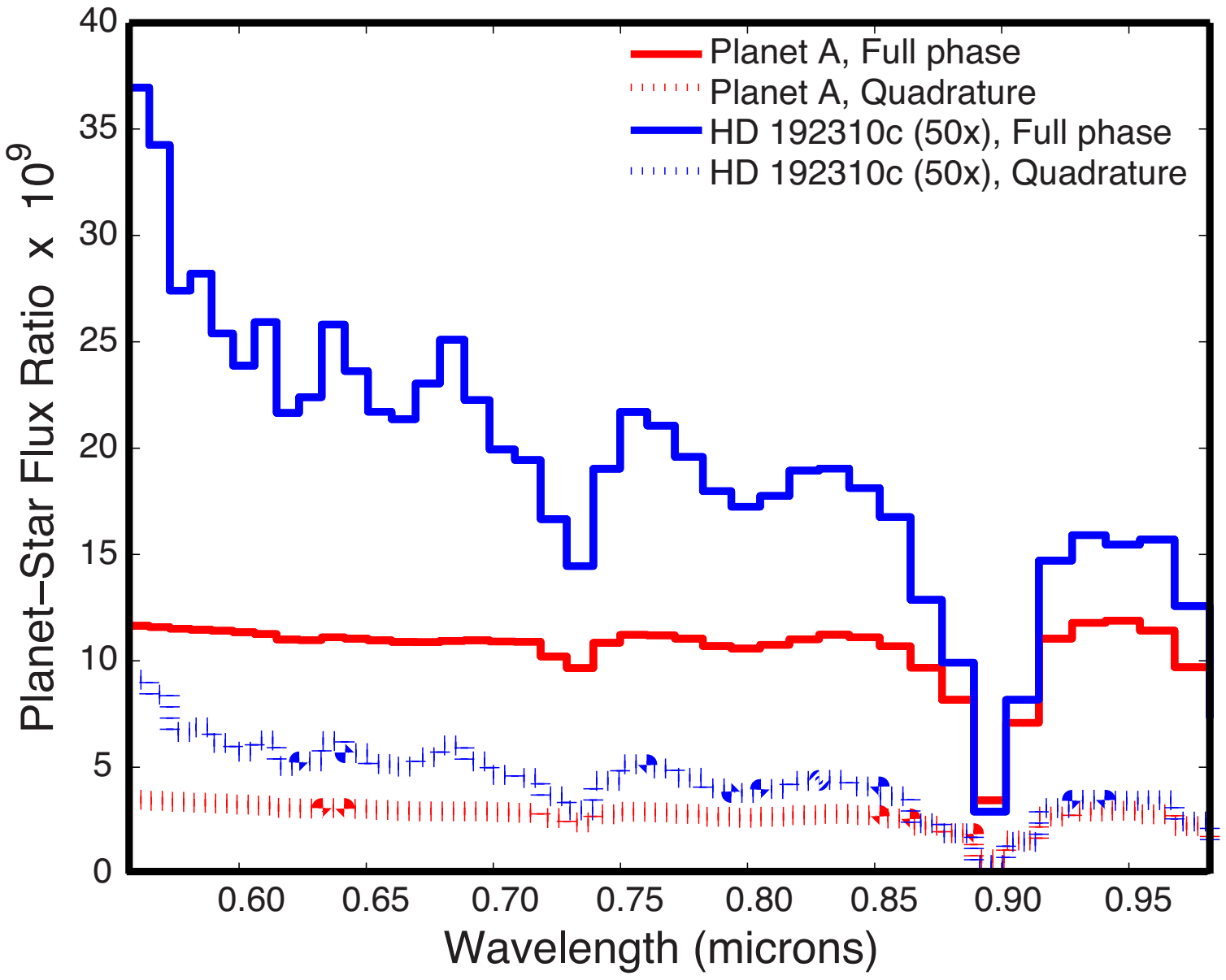

**Figure 4. Model noise-free contrast spectra for the two test cases, Planet A (red) and HD 192310c (50x metallicity, blue), for a spectral resolution of R = 70. Dashed spectra indicate an observation at quadrature; methane absorption features are still notable.**

## 3. Results

### 3.1 Retrieved Best-Fit Spectra by MCMC

We generate model albedo spectra for Planet A (Figure 5) and HD 192310c (Figure 6) at varying phase angles similar to Figure 4; in this study we explore phase angles of 30°, 40°, 60°, 70°, 90° and 120°. Both resulting spectra are then combined with an instrument-specific model, in this case, the notional WFIRST noise model (Robinson et al. 2015; Table 2).

While ***Paper 1*** performs retrievals against an albedo spectrum, since this work is also concerned with planet radius, we retrieve against the planet-to-star flux ratio, or contrast spectrum. Contrast spectra are created for a range of signal-to-noise ratios (SNR) to explore observational limits and their effect on observations. Here we define the SNR to be centered at $\lambda = 0.6$ micron with an 8.6 nm wide bandpass. SNRs of 5, 10 and 20 are explored, as in ***Paper 1***. To be clear, since SNR is wavelength-dependent, these values refer to the SNR at 0.6 $\mu$m.

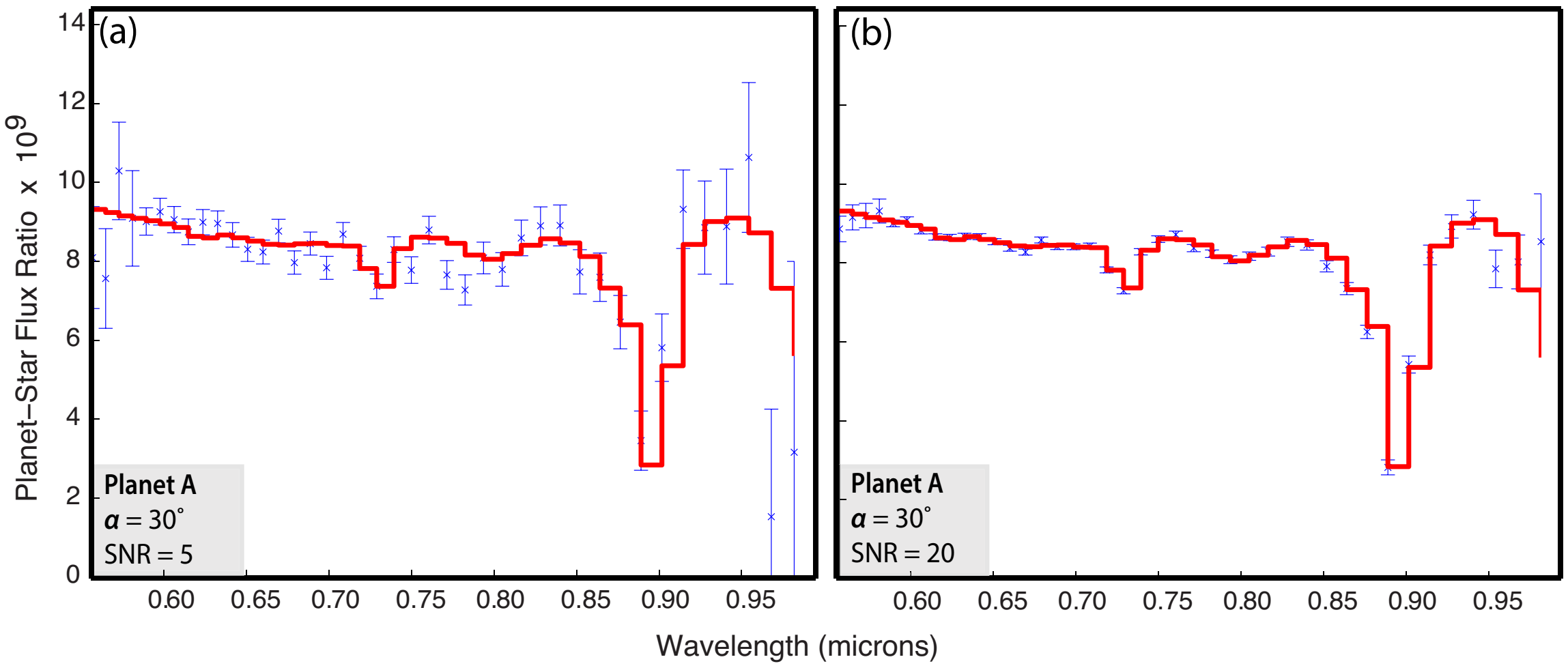

**Figure 5. Contrast spectra for Planet A at SNR values of (a) 5 and (b) 20. Spectra are generated at a spectral resolution of R = 70 and a phase angle of 30°. Red represents the truth spectra and blue error bars represent notional instrument noise during observation** (Robinson et al. 2015)**.**

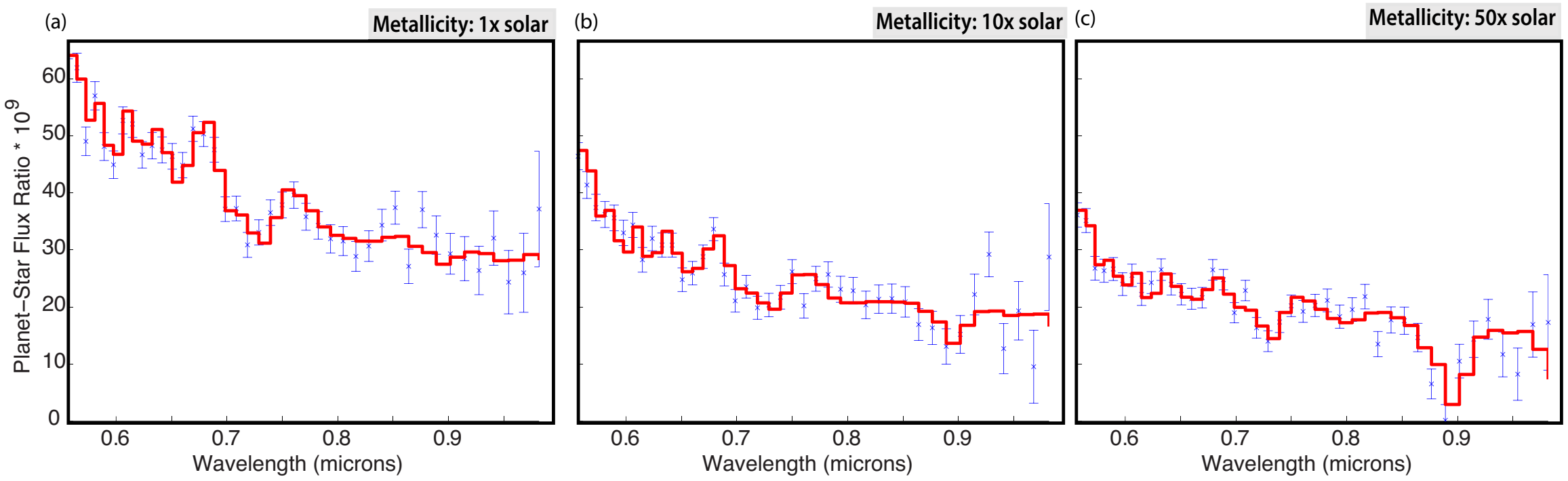

**Figure 6. Contrast spectra for HD 192310c generated at metallicity values of (a) 1x, (b) 10x and (c) 50x that of the Sun. The difference in the methane absorption signature at ~0.9 micron is evident. Spectra are generated at a phase angle of 0°, SNR = 20 and a spectral resolution of R = 70.**

For each of the eleven retrievable parameters (Table 1), we employ twelve MCMC chains, or "walkers", for a total of 132 chains. Each chain was then run for 2500 iterations for a final sample chain of 330,000 samples. Selected cases were run for 4000 iterations to ensure that the MCMC algorithm did not get stuck in local minima and was exploring the entire parameter space; returned ranges were found to be nearly identical to the 2500 iteration run, so we restrict ourselves to 2500 iterations for all the results presented here. We identified the best fit, 1σ and 2σ-range of retrieved spectra for Planet A; the retrieved models match the "true" spectra well. Similar excellent fits are seen for phase angles between 30° and 120° despite decreasing contrast signals at larger phase angles (Figure 7).

We perform a similar study against HD 192310c using the same MCMC parameters as for Planet A. For this planet, we generate three separate test cases by constructing forward models and performing retrievals at three different metallicities, namely, one, ten and fifty times solar values (1x, 10x, 50x). For each metallicity case, we generate model albedo spectra at 30°, 40°, 60°, 70°, 90° and 120° phase angle and apply the notional WFIRST noise model to them.

The resulting spectra reveal a variety of methane absorption signatures (Figure 6). Notably, because of the relatively high cloud, the 1x solar case exhibits particularly subdued methane absorption features. Figure 8 - Figure 10 illustrate the corresponding spectral recoveries across three values of signal-to-noise (SNR) and three values of metallicity, at phase angles of 30° and 90°.

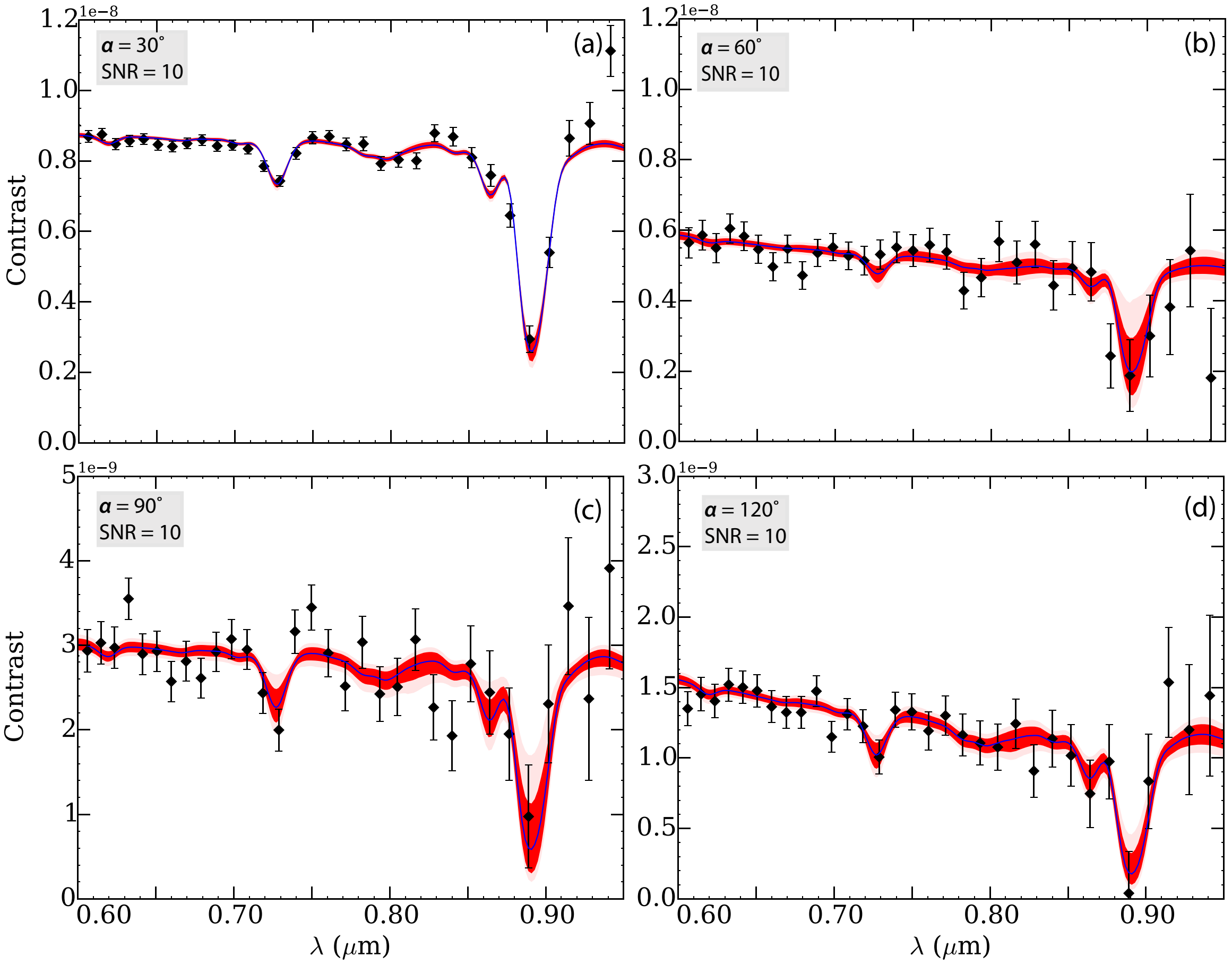

**Figure 7. Best-fit contrast spectra for Planet A at SNR = 10 and varying phase angles: (a) 30°; (b) 60°; (c) 90°; (d) 120°. Model spectra are calculated from the best 198,000 samples (1500 iterations) of the 330,000 final sample chain (2500 iterations). The median spectrum (blue) matches well to the truth**

**spectrum (black). 16-84% (dark red) and 4.5-95.5% (light red) percentile range of recovered solutions are also shown. Good matches to the model truth spectra are seen in all cases, even at relatively low contrast signals for large phase angles. Note the differing (smaller) vertical scales between (a) and (d).**

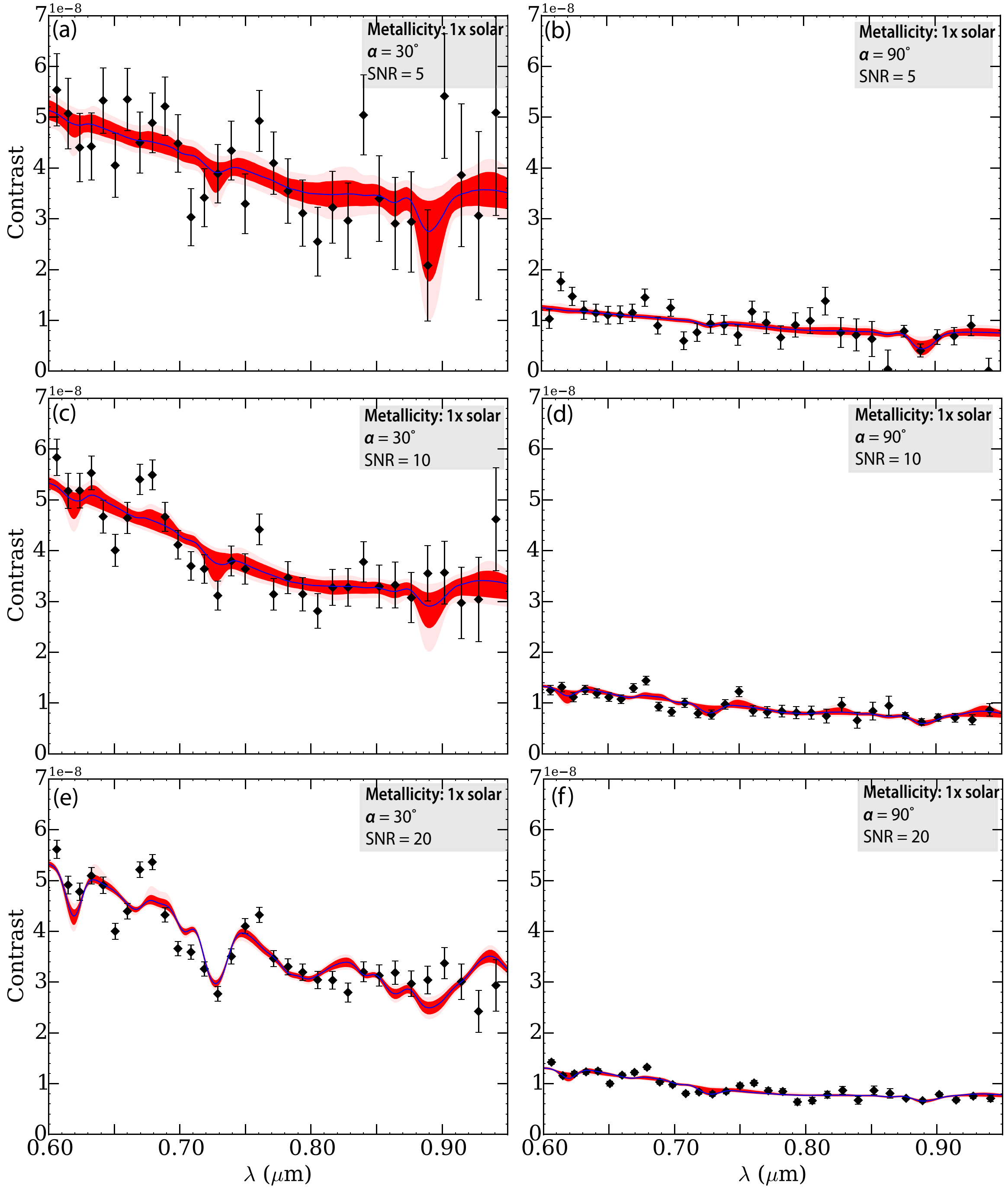

**Figure 8. Best-fit contrast spectra for HD 192310c at a metallicity of 1x solar and varying SNR and phase angle: (a) SNR = 5, α = 30°; (b) SNR = 5, α = 90°; (c) SNR = 10, α = 30°; (d) SNR = 10, α = 90°; (e) SNR = 20, α = 30°; (f) SNR = 20, α = 90°. Note the lack of the characteristic methane absorption signal at 0.9 microns for the 90° case as compared to the 30°. Description of colors is as in Figure 7.**

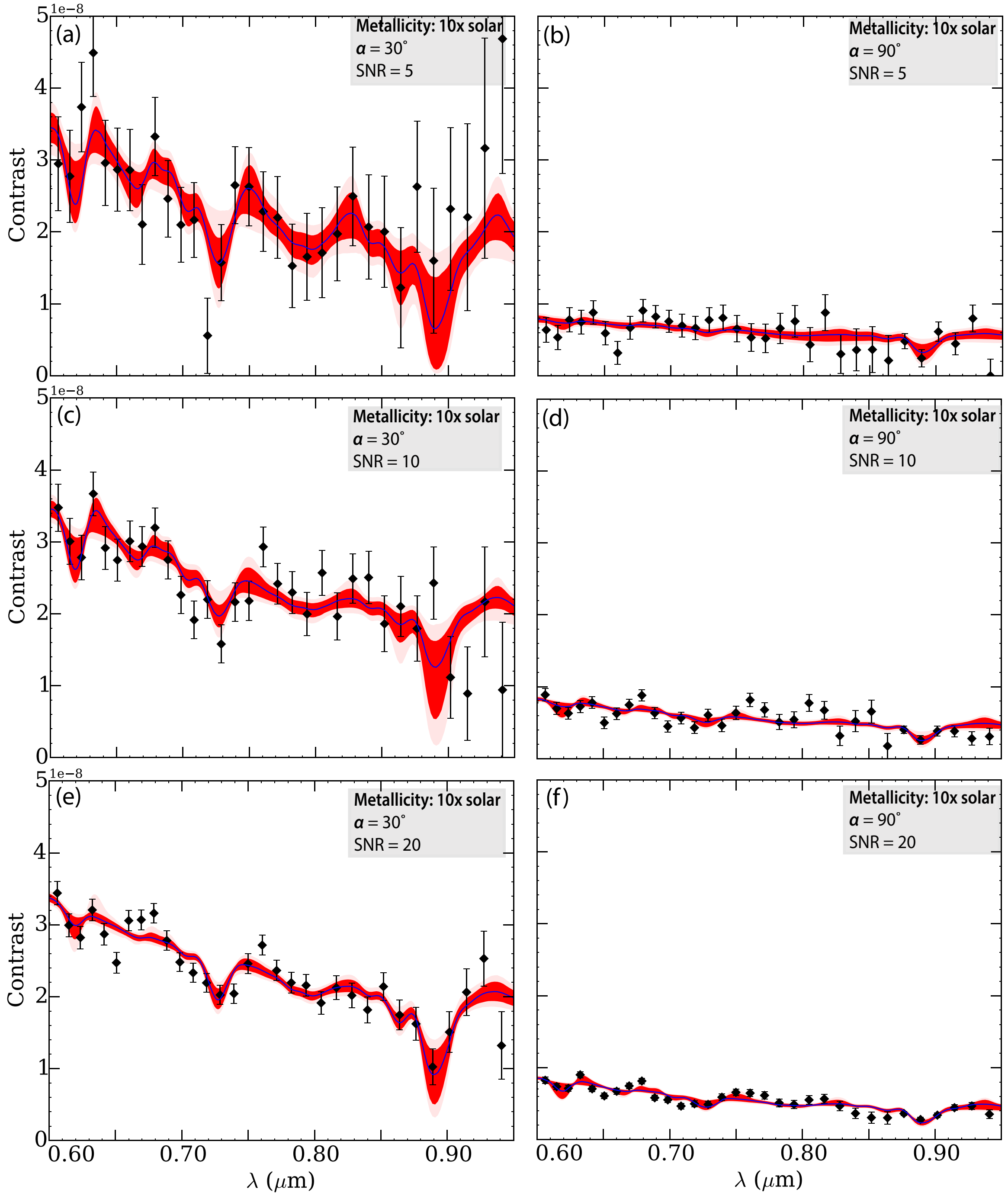

**Figure 9. Best-fit contrast spectra for HD 192310c at a metallicity of 10x solar and varying SNR and phase angle: (a) SNR = 5, α = 30°; (b) SNR = 5, α = 90°; (c) SNR = 10, α = 30°; (d) SNR = 10, α = 90°; (e) SNR = 20, α = 30°; (f) SNR = 20, α = 90°. Good matches to the model truth spectra are seen in all cases, even at relatively low contrast signals for large phase angles. The improvement in recovered signal is evident with increasing SNR from (a) through (f). Description of colors is as in Figure 7.**

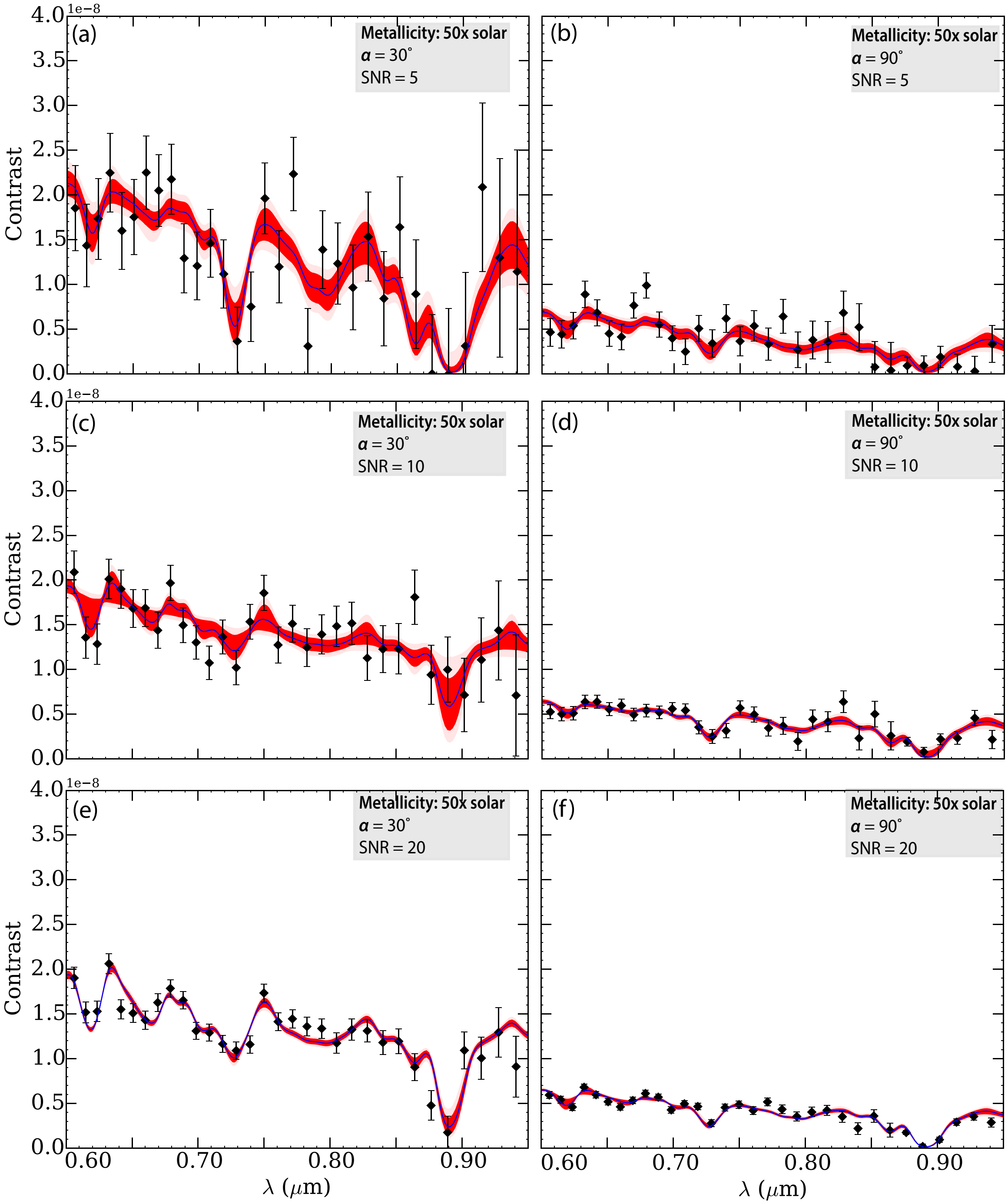

**Figure 10. Best-fit contrast spectra for HD 192310c at a metallicity of 50x solar and varying SNR and phase angle: (a) SNR = 5, α = 30°; (b) SNR = 5, α = 90°; (c) SNR = 10, α = 30°; (d) SNR = 10, α = 90°; (e) SNR = 20, α = 30°; (f) SNR = 20, α = 90°. Good matches to the model truth spectra are seen in all cases, even at relatively low contrast signals for large phase angles. The improvement in recovered signal is evident with increasing SNR from (a) through (f). Description of colors is as in Figure 7.**

## 3.2 Inferring Phase-dependent Relationships from Posterior Probability Distributions

We assemble retrieval results similar to those shown in Section 3.1 into posterior probability plots, which graphically show the marginal probability distribution between every retrieved parameter pair. Figure 11 is an example of this plot. Here, darker colors represent a higher probability of the solution lying in that region, and the diagonal of the plot shows the marginalized probability distribution, to a 68% confidence interval, for each retrievable parameter in Table 1.

A broad distribution means that the parameter is largely unconstrained, as is the case for phase angle (Figure 11, marker a) or surface gravity. Conversely, a sharp peak in the distribution and small ranges on returned values means that the parameter can be well determined, such as planet radius (Figure 11, marker b). In other cases, more general relationships can be inferred, for example, lower limits to the albedo of the top cloud (Figure 11, marker c) and methane abundance. Such plots illuminate how variations in retrieved values vary with changes in the "true" planet phase, as well as interrelationships between other parameters. A brief discussion of general trends apparent in the relationships of phase angle with other retrievable parameters follows; though we use the Planet A case (Figure 11) to highlight these trends, they are also seen for HD 192310c.

Figure 12 highlights relationships between key parameters from a sample probability distribution. First, the relationship between radius and phase is in line with our conceptual understanding from Figure 1: for larger phase angles (i.e. a more crescent phase), a larger planet radius is favored. Even at a relatively high phase angle (60°) a clear detection of methane, with a lower limit to the atmospheric mixing ratio of larger than $10^{-3.5}$, is seen. Surface gravity appears essentially unconstrained, although as we will show later, a larger SNR or smaller phase angle does narrow the probable range. The difficulty in deriving meaningful constraints on gravity from reflection spectra is discussed in more detail in ***Paper 1***.

The MCMC analysis appears to constrain the top cloud well (quantity $dP_2$, see Table 1 and Figure 2), but is indeterminate on the pressure “gap” between the cloud layers (quantity $dP_1$). The pressure of the bottom cloud (quantity P) is not tightly constrained, but higher probability values are distributed around the true value. This could imply either that the bottom cloud is not well constrained, or perhaps that a one-layer cloud model is more suitable for this planet. Previous MCMC simulations on HD 99492c reach a similar conclusion (Marley et al. 2014).

We generate marginalized probability distributions at seven phase angles (0° - 120°) and three SNR values (5, 10, 20). For the case of SNR = 20, Figure 13 shows the relationship between retrieved phase angle and planet radius for phase angles from 30° to 120°. While the error bars on retrieved planet phase angle are large, at both low and high phase angles, the MCMC algorithm retrieves best-fit values close to the true value. However, at phase angles between 45° and 90°, the probable phase angle values stretch across most of the phase angle solution space, making it more difficult to obtain good values. Observations with comparable SNR and multiple phase angles can therefore be extremely valuable in determining both orbital characteristics, if unknown or uncertain, and narrowing down planet radius.

Finally, we collate summary plots of the retrieved parameters at all seven phase angles (0° - 120°) and SNR values (5, 10, 20) for all four test cases. The retrieved values of methane abundance, surface gravity, planet radius, recovered phase angle and cloud pressures are respectively plotted against changing SNR and phase angle for Planet A (Figure 14), HD 192310c 1x (Figure 15), 10x (Figure 16) and 50x (Figure 17) cases. Colors represent the size of the 68% confidence interval values, seen for each parameter on the diagonal of probability distribution plots such as Figure 11.

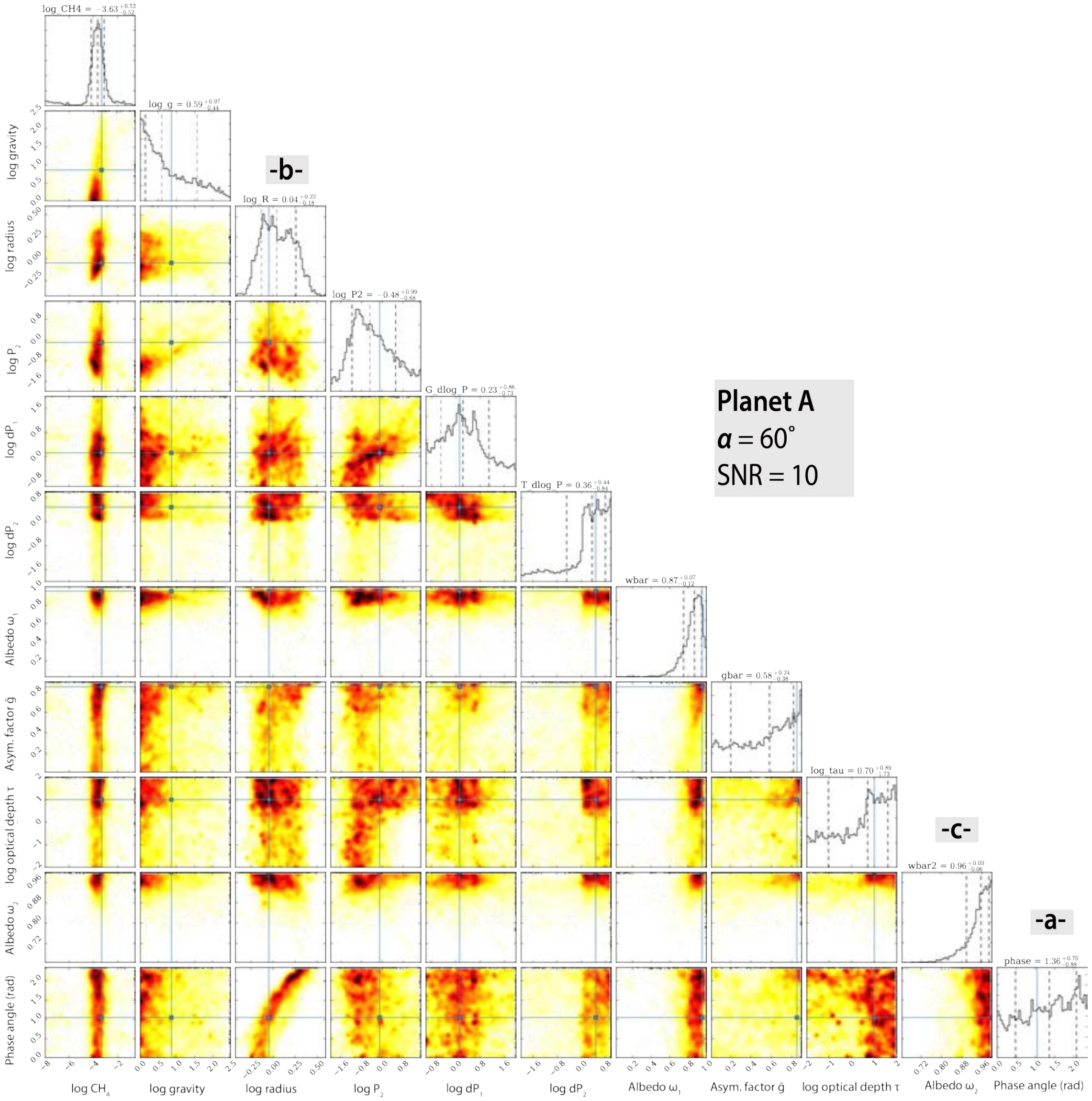

**Figure 11. Posterior probability distribution plot for all eleven parameters retrieved by the MCMC algorithm (Table 1), for the case of Planet A at SNR = 10 and phase angle 60° (Figure 7c). Darker regions represent higher probability. Blue dots represent true values. The distributions are drawn from all remaining samples after the MCMC burn-in chains (first 1000 chains) are discarded. The diagonal of the plot represents the marginalized probability distributions for each parameter. Best-fit values in log space (except for phase angle) are shown. See text for references to text markers a-c. The error bars indicate (left to right) the 16%, 84% and 50% quantiles, i.e., this is the 68% confidence interval.**

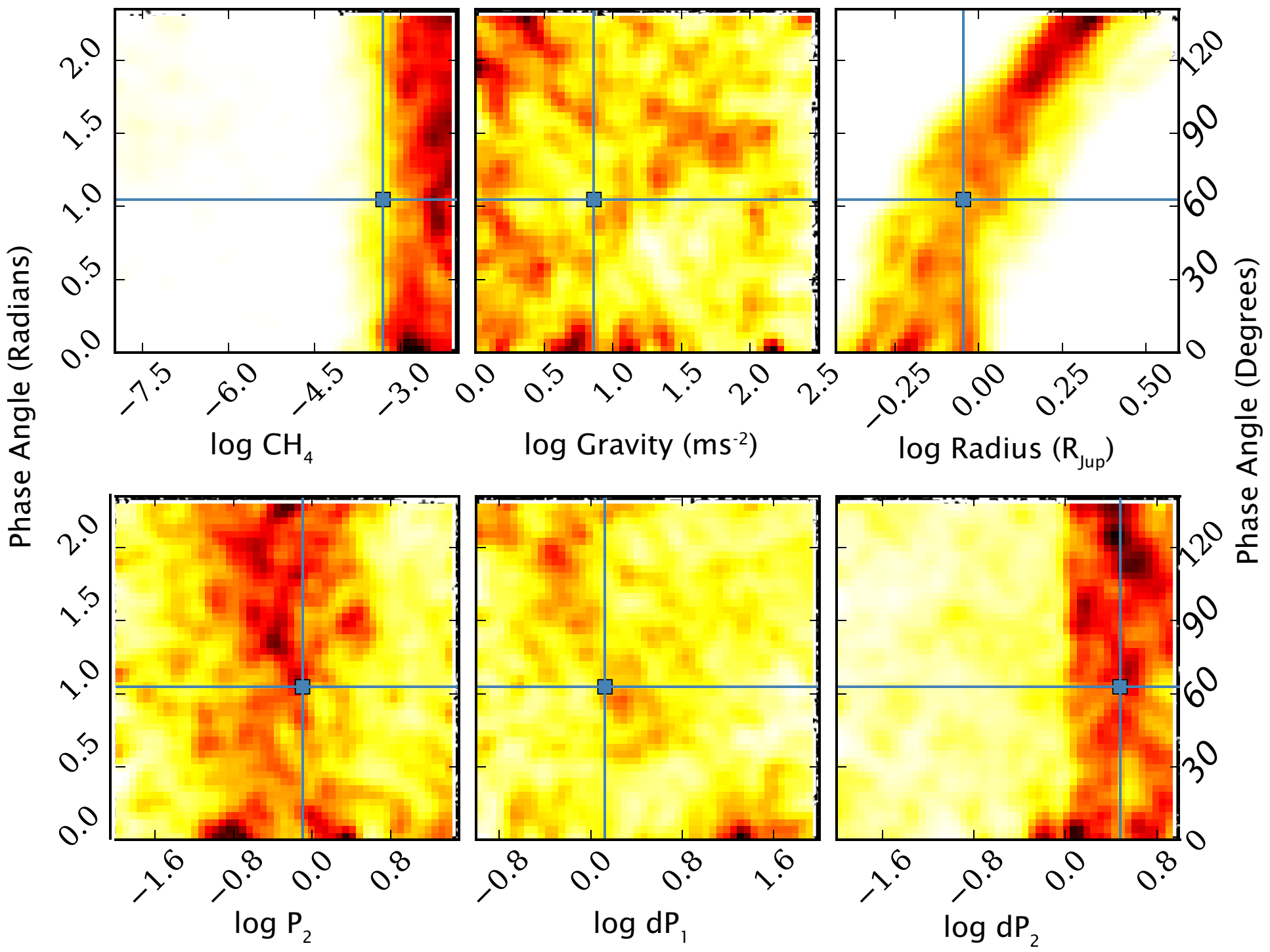

**Figure 12. Sample highlight of pertinent parameter relationships with planet phase angle. The top row shows probability distributions of phase angle against methane abundance, surface gravity and planet radius. The bottom row shows distributions of phase angle against, respectively, the pressure of the bottom cloud (P) in the two-layer model by Marley et al. (2014), the pressure difference between the two cloud layers ($dP_1$) and the pressure difference across the top cloud layer ($dP_2$). All parameters are in log space with the exception of phase angle.**

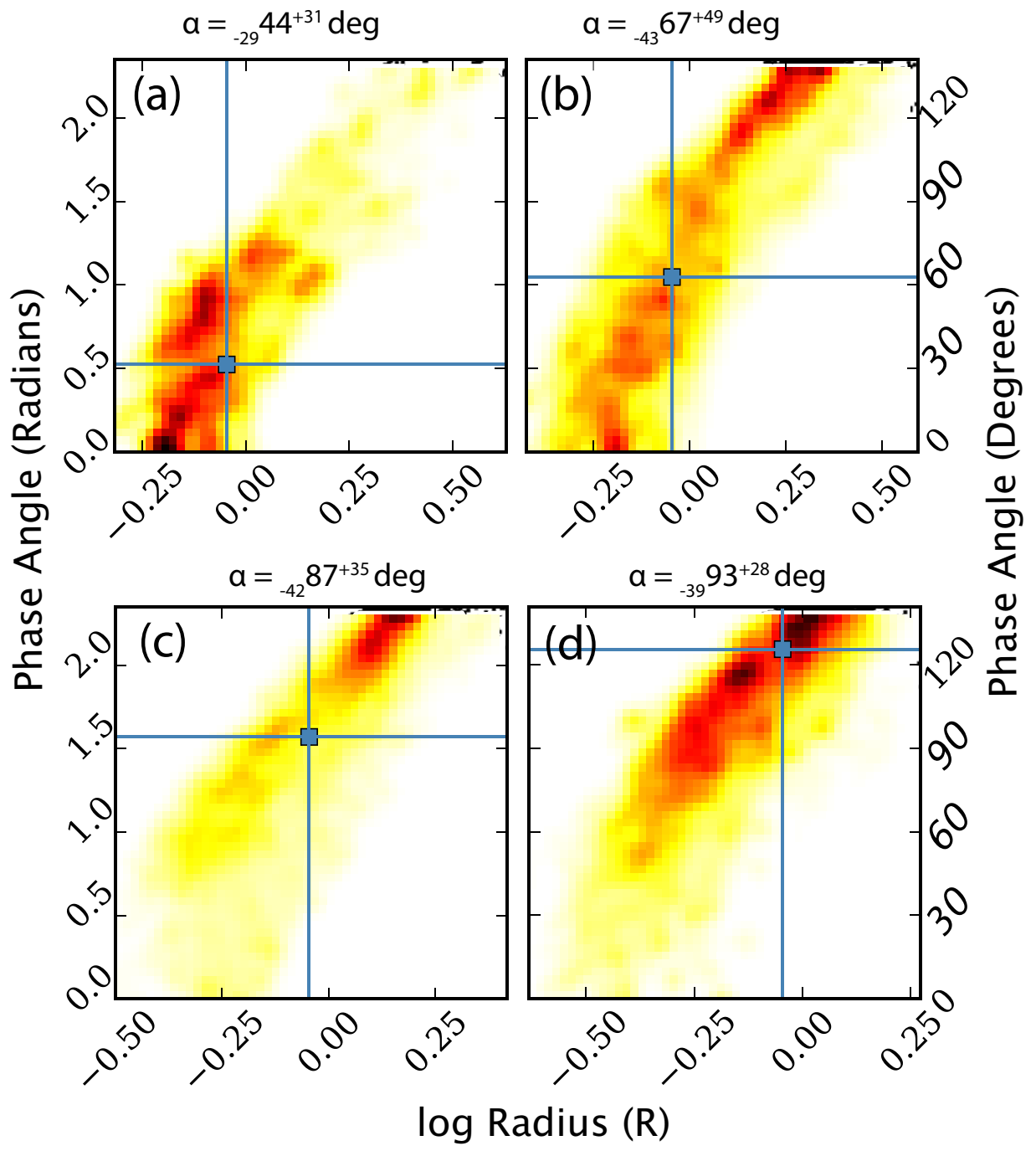

**Figure 13. The relationship of planet radius with changing planet phase angle, for the Planet A test case, for an SNR of 20 and a truth phase angle (blue square) of: (a) 30°; (b) 60°, (c) 90° and (d) 120°. Best-fit values for each case are indicated above the figure; superscripts and subscripts to this value indicate**

upper and lower bounds returned from the posterior probability plot diagonals (e.g. Figure 11). The MCMC algorithm retrieves phase angle and radius values close to the true value for planet phases close to full phase and past quadrature.

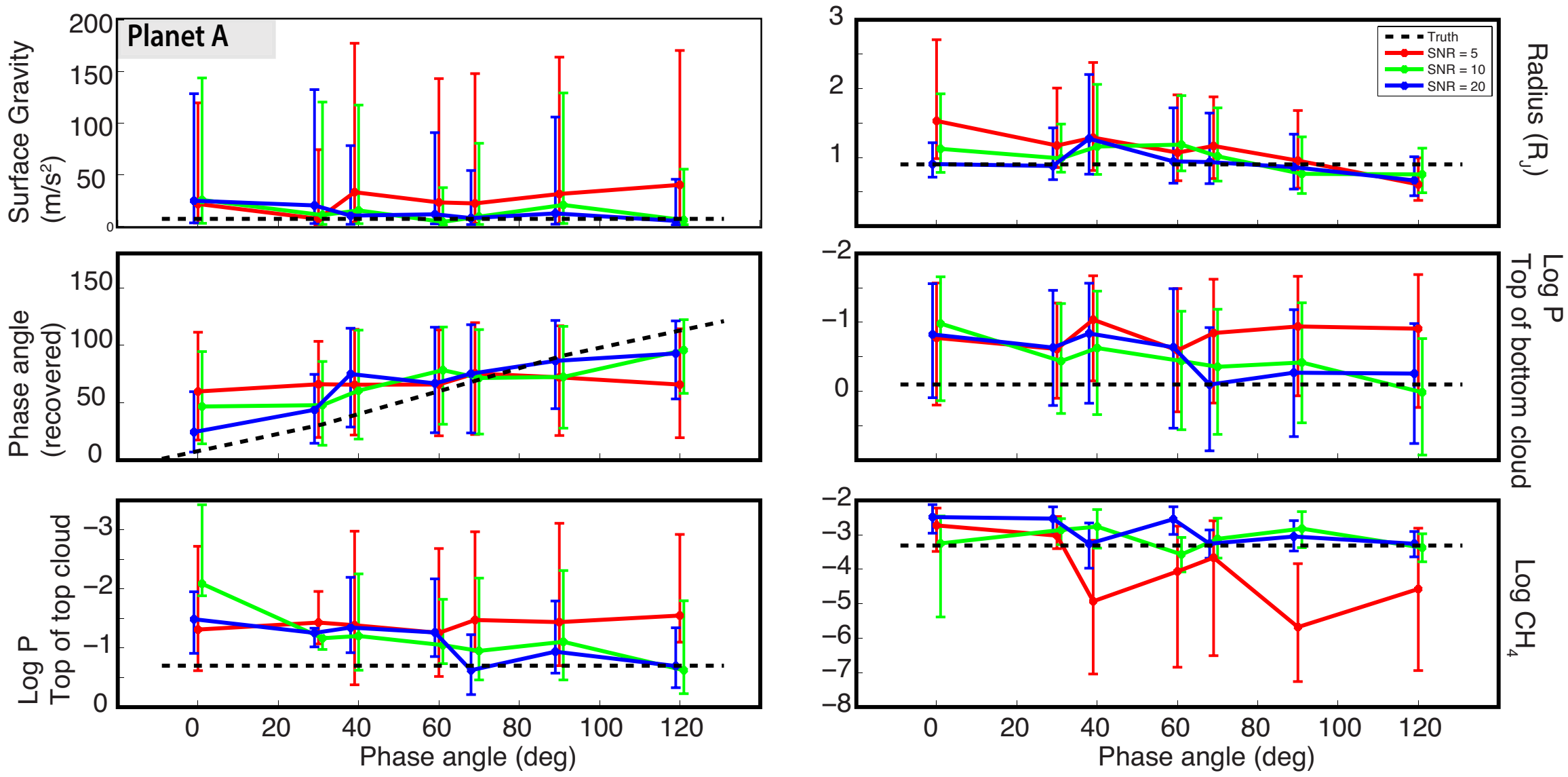

**Figure 14. Summary of all results for Planet A test case. True phase angle varies from 0° to 120°. SNR varies between 5 (red), 10 (green) and 20 (blue). Error bars enclose the 68% confidence interval as defined by MCMC posterior probability distributions similar to Figure 11. Solid dots denote the best-fit values. A black dashed line denotes true values from the Planet A model.**

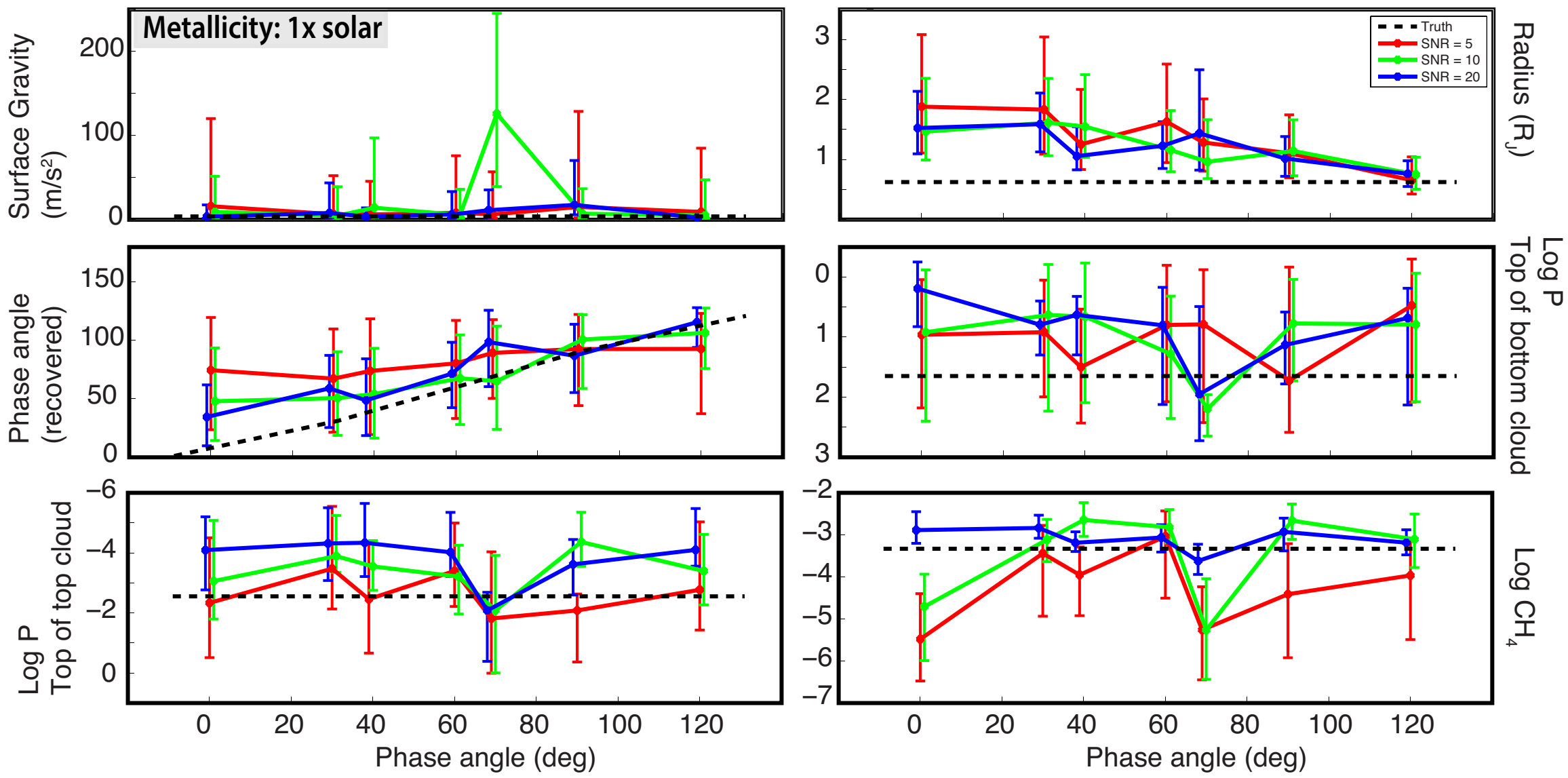

**Figure 15. Summary of all results for the HD 192310c test case with 1x metallicity of the Sun. True phase angle varies from 0° to 120°. SNR varies between 5 (red), 10 (green) and 20 (blue). Error bars are as in Figure 14. Solid dots denote the best-fit values. A black dashed line denotes true values from the HD 192310c model.**

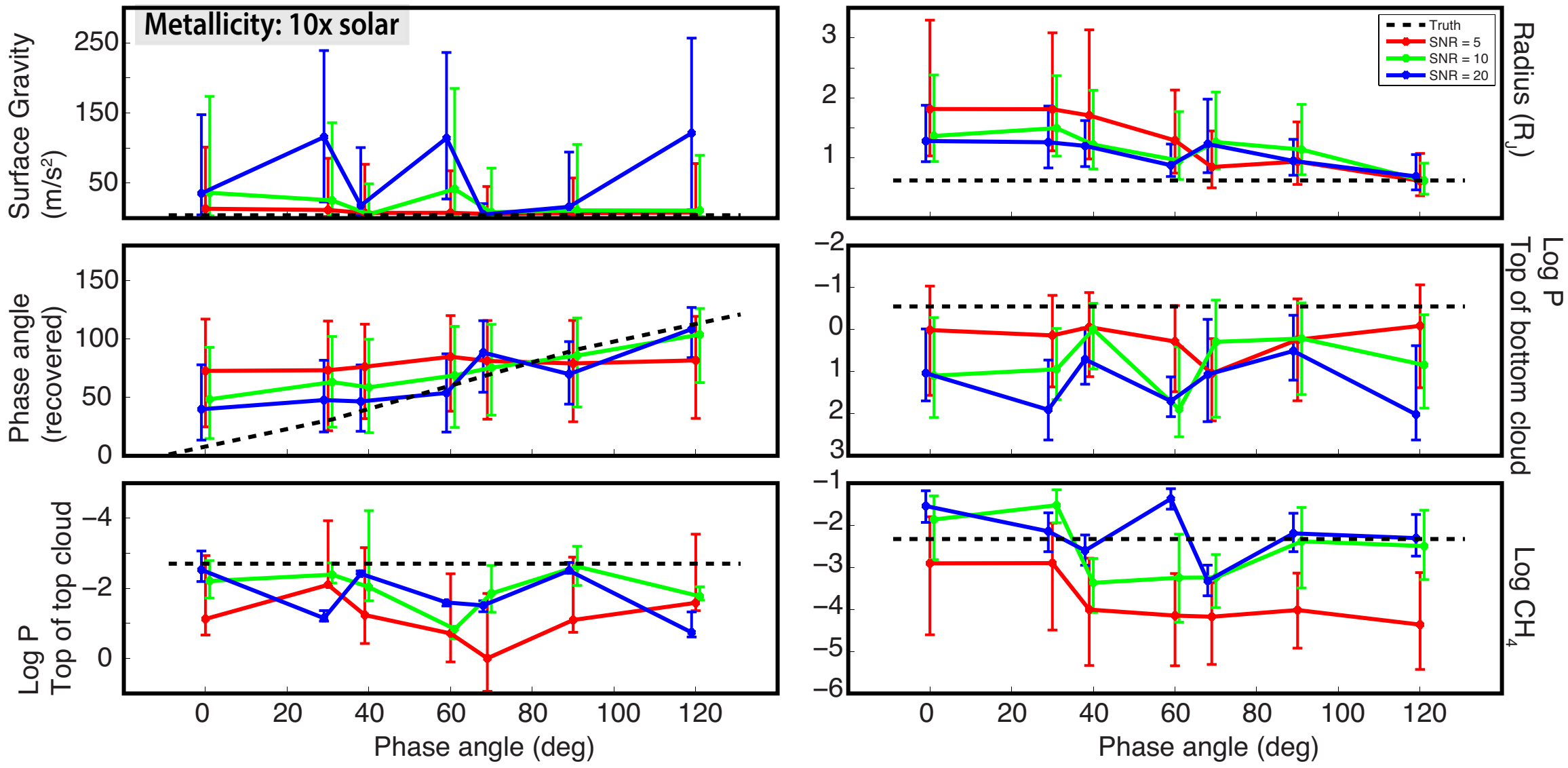

**Figure 16. Summary of all results for the HD 192310c test case with 10x metallicity of the Sun. True phase angle varies from 0° to 120°. SNR varies between 5 (red), 10 (green) and 20 (blue). Error bars and solid dots are as in Figure 14. A black dashed line denotes true values from the HD 192310c model.**

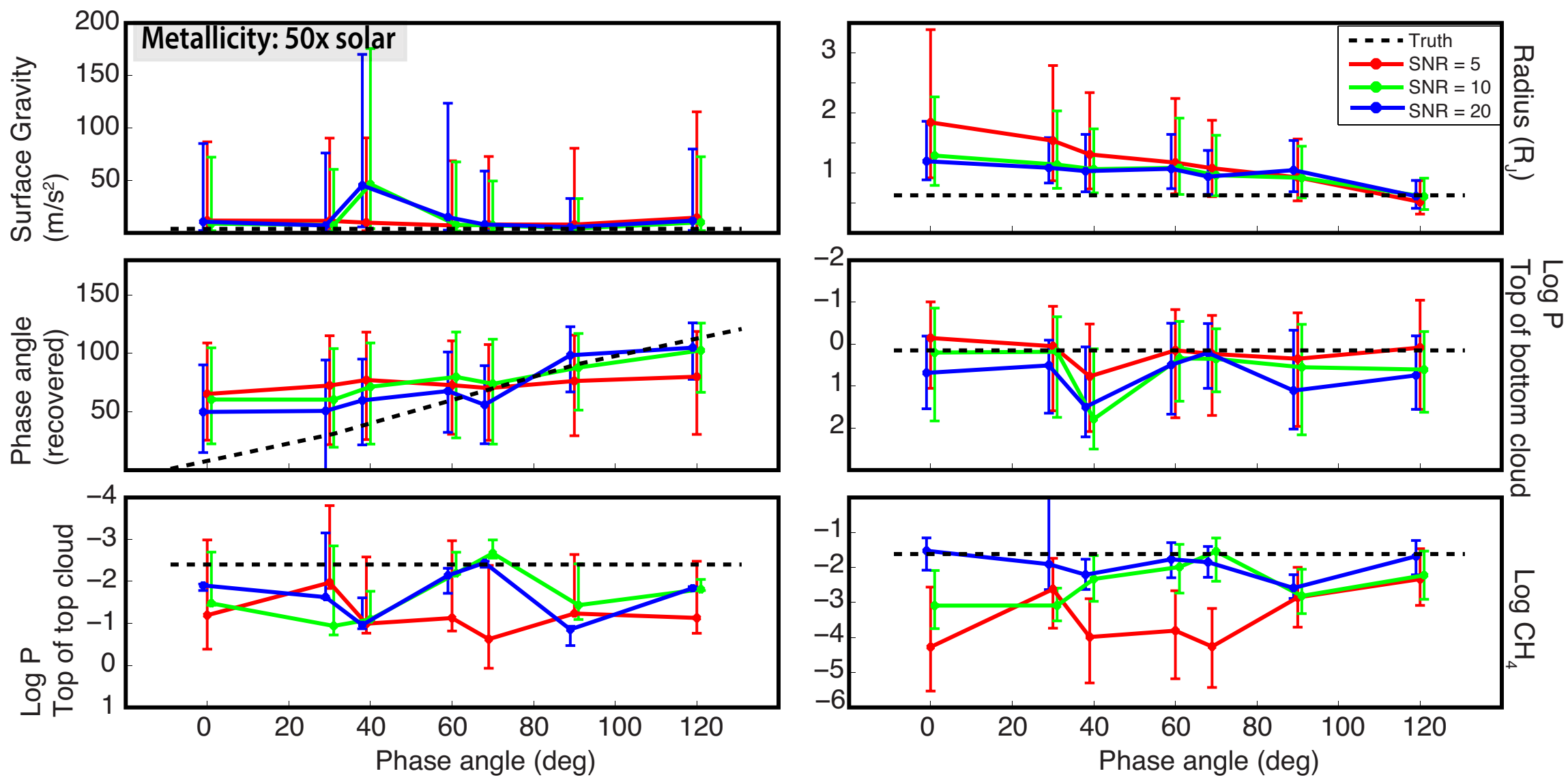

**Figure 17. Summary of all results for the HD 192310c test case with 50x metallicity of the Sun. True phase angle varies from 0° to 120°. SNR varies between 5 (red), 10 (green) and 20 (blue). Error bars and solid dots are as in Figure 14. A black dashed line denotes true values from the HD 192310c model.**

Several trends are apparent from the summary figures. Generally speaking, we find that retrievals at higher SNR ratios (SNR = 20) place correct constraints on the atmospheric methane abundance, to within an order of magnitude. The low SNR=5 case identifies the presence of methane, but the abundance is highly uncertain (four orders of magnitude), as many combinations of cloud top pressure, phase angle, and gravity are able to adequately reproduce the noisy data.

In general, the high bright clouds found in this case seem to lead the retrievals to favor cloud tops deeper in the atmosphere than in the forward model, with lower brightness compensated by larger planetary radii. This is seen for all metallicities, particularly at low phase angles. The 1x metallicity case with the weakest methane features clearly presents a particular challenge, even at SNR of 10, as the methane abundance is nearly unconstrained. For all three metallicities considered, only the case with an SNR of 20 systematically reproduces close to the correct methane abundance at all phase angles. We discuss these findings further in the next section.

## 4. Discussion

In this section, we discuss how well the planet radius, phase angle, methane abundance and cloud properties were constrained in the presence of planet phase and radius uncertainties. We focus the discussion in this paper on the newly introduced uncertain phase and radius determinations, as the abundance and cloud properties were the focus of ***Paper 1***.

### 4.1 Methane and Radius Retrieval

For the cases considered here we assumed that phase angle was almost completely unconstrained. While this is a situation unlikely to be encountered for a real planet, it is a difficult bounding case worthy of additional study. We find that phase angle is generally not well constrained from a single observed spectrum. Since changing planetary parameters, including planet radius, cloud height or phase angle can all create degenerate changes to reflected light flux, large uncertainties can result in all these parameters. Generally speaking, with no prior knowledge of orbital parameters from radial velocity, the most we can confidently tell about the phase angle from retrievals is whether it is *high* or *low* (above or below ~90°).

One might expect the planet radius solution space to be similarly large, but this is not the case. Despite being given an impossibly large range of 0.1 – 100 $R_J$, even at a low SNR of 5, the MCMC routine typically returns a solution within a factor of two of the true value (~1 $R_J$). Regardless of true phase angle, the MCMC algorithm must match the observed flux from the planet. At more crescent phases, the atmosphere is highly forward scattering and molecular bands are weak. Since clouds are relatively less important at such scattering angles, this drastically reduces the number of free parameters. Consequently, we find the most accurate radius retrievals at the highest phase angles.

### 4.2 The impact of a known phase angle

Given the difficulty in retrieving the true phase angle from a single spectrum, we also explored the impact of a better-constrained phase angle, as might be expected during an observational campaign. To study this case we chose the 10x metallicity case for HD 192310c, at a favorable SNR of 20. Retrievals were performed on this case at multiple phase angles, given a $\pm 10°$ restriction from the true value, on the possible values of the angle.

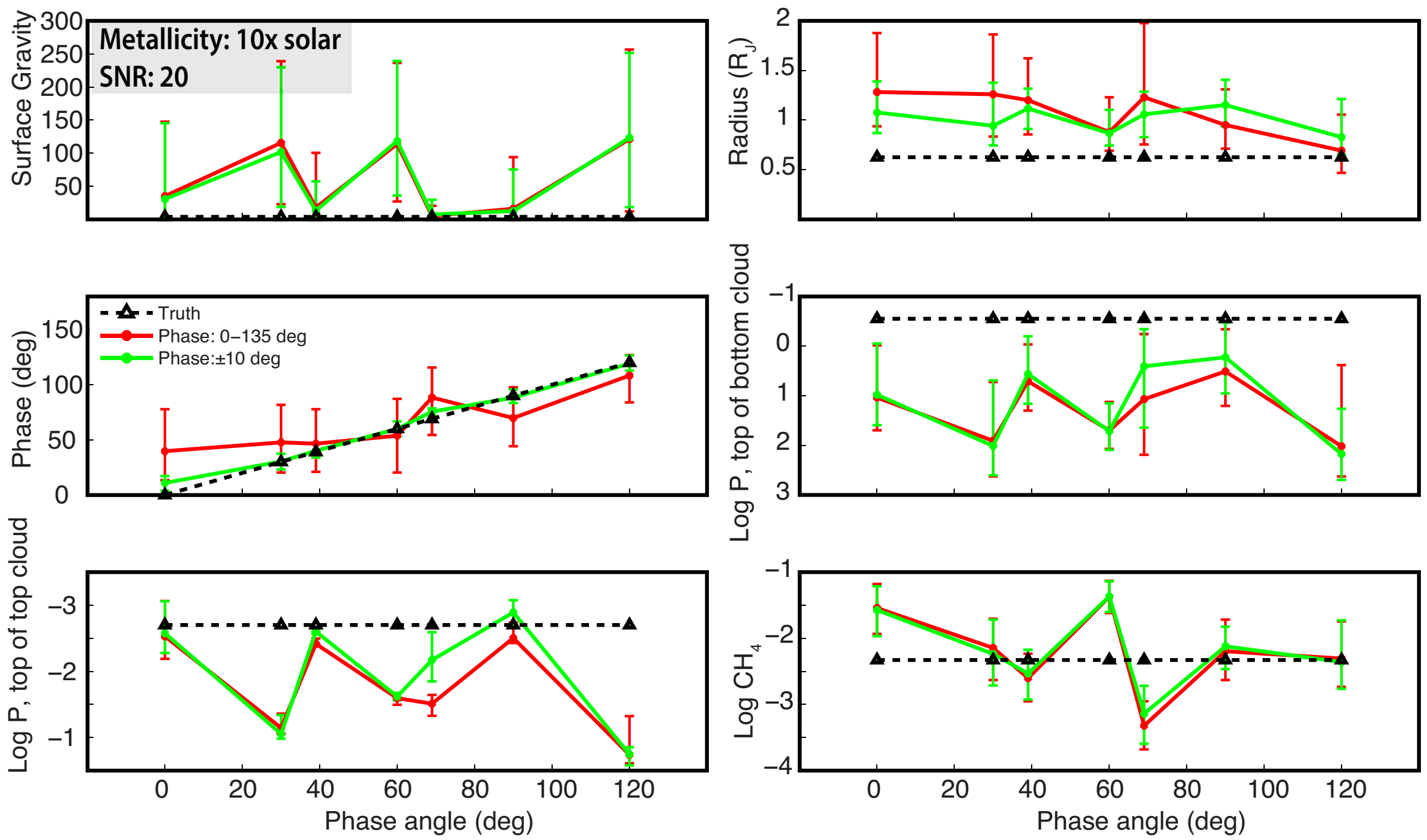

**Figure 18. Results for the 10x metallicity case of HD 192310c, at SNR = 20, for an unbounded phase angle case (red) and a bounded case (green), where for the bounded case, the phase angle can vary by no more than 10° from the true value. No impact to cloud property or methane abundance retrievals are seen, however, the retrieved radius of the planet improves by a factor of two.**

As seen in Figure 18, this does not significantly improve retrievals of gravity, cloud properties or the methane abundance. The only noticeable difference is that phase angle knowledge helps constrain the radius of the planet better, improving the radius determination by a factor of two. Given a proxy value for planet mass from radial velocity measurements ($M$ sin $i$), by improving knowledge of the planet radius, prior knowledge of the phase angle best helps improve the estimate of surface gravity ($M/r^2$), though this is not seen directly from gravity retrievals.

## 4.3 Applying an intersection criterion to multiple observations

It has been shown that planet radius retrievals, for example, improve with increasing phase angle, whereas retrievals for top cloud pressure improve with decreasing phase angle. Bounds on quantities such as methane abundance and cloud properties vary significantly with SNR. Simultaneous retrievals on observations taken at multiple phase angles would therefore likely hold promise for narrowing the solution space of best-fit models. While we did not perform simultaneous retrievals, we present a preliminary investigation into their utility by imposing an intersection criterion.

The intersection criterion is defined as in set theory: for sets $A$ and $B$, the intersection of sets ($A \cap B$) defines that set which contains only elements of $A$ that also belong to $B$. For two observations taken at different phase angles, and separate retrievals, applying an intersection criterion means only those solutions that appear in both retrievals are considered valid. Here we use "observation" to refer to the integration time needed to produce one complete ~600-970 nm spectrum, at a given SNR. The intersection criterion is illustrated in Figure 19. While clearly an estimation, simultaneous retrievals against combined phase-varying datasets are planned as future work.

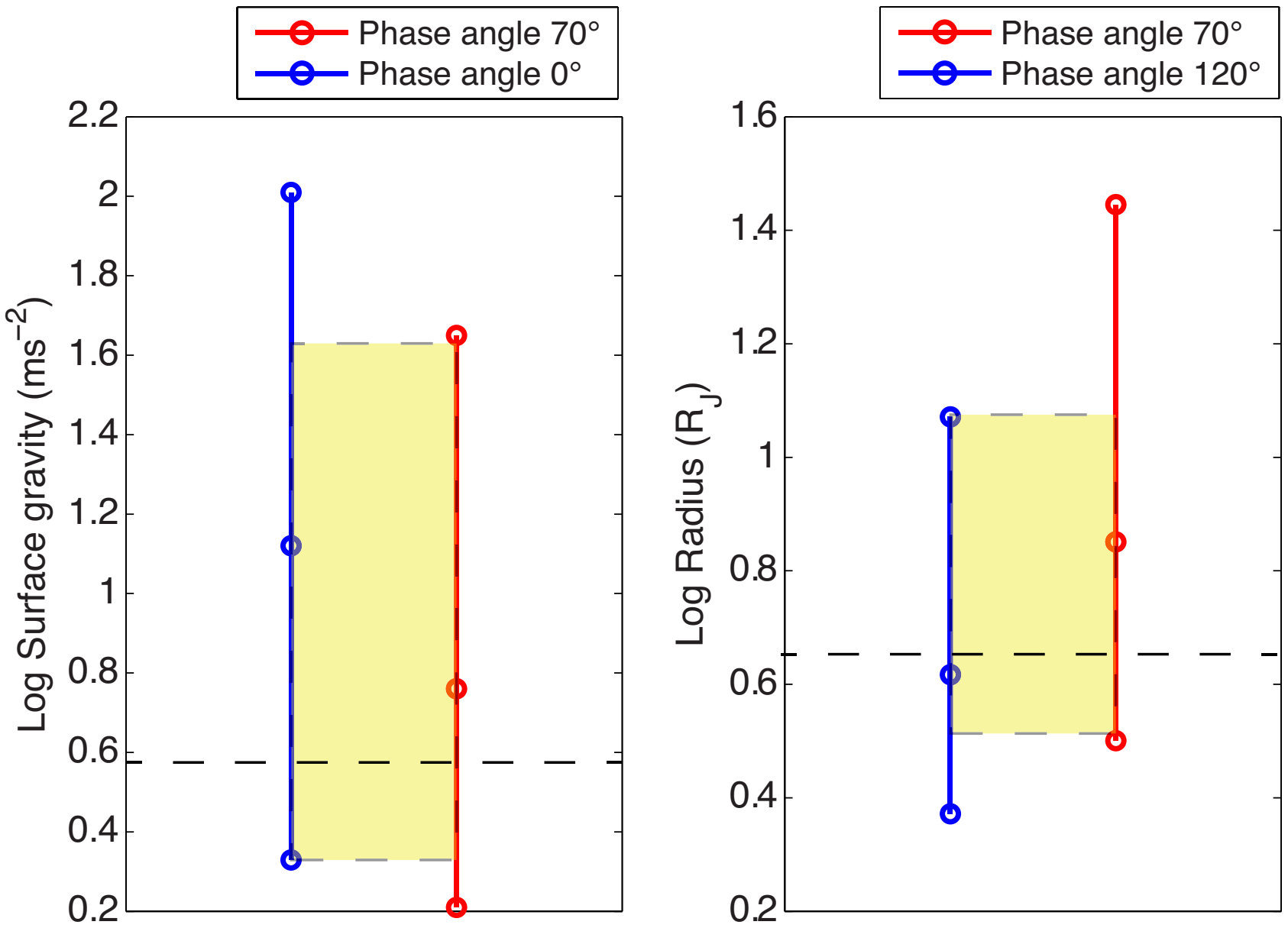

**Figure 19. Illustration of the intersection criterion for surface gravity and planet radius (note log units for both). Data is from retrievals for HD 192310c, metallicity 10x solar and an SNR of 5. Phase angles are shown in the legend. The dashed black line indicates true values. The highlighted yellow area represents the region of solutions common to both retrievals (intersection criterion); it can be seen to visibly improve error bars on both cumulative solutions.**

The underlying idea is that intersection of multiple observations at varying phase angles may determine a more likely range for parameters of interest. We begin by determining which combination of phase angles will be likely to improve retrievals the most. This analysis focuses on the lowest SNR case, as the uncertainty on retrieved results with one set of observed spectra is the highest, and multiple observations are likely to have the most impact. The HD 192310c 10x case is chosen again here, although a similar analysis may be conducted for any planet targeted as part of an observational campaign.

Figure 20 shows the improvement in 68% confidence intervals achieved by applying the intersection criterion to all phase angle solutions, for HD 192310c 10x, at an SNR of 5. Here, "improvement" denotes the difference between *1)* confidence intervals obtained using single observations and *2)* intervals obtained by applying an intersection criterion to multiple observations taken at differing phase angles. A value of zero for improvement represents one of two cases; either *1)* there is no overlap between retrieved values at different phases, and the solutions must be considered independently, or *2)* returned solutions are completely identical at both phase angles. In either case, considering multiple observations does not represent an improvement over a single observation. Conversely, peaks represent cases where confidence intervals were tightened by applying an intersection criterion to multiple observations, i.e., we can estimate which combination of phase angles may be most helpful to reduce uncertainty in retrieved quantities. For example, for methane, if a first observation is taken at $\alpha = 70°$ (x-axis), a subsequent observation at $\alpha = 30°$ (yellow line) would improve the 68% confidence interval by approximately 1.4 orders of magnitude (y-axis).

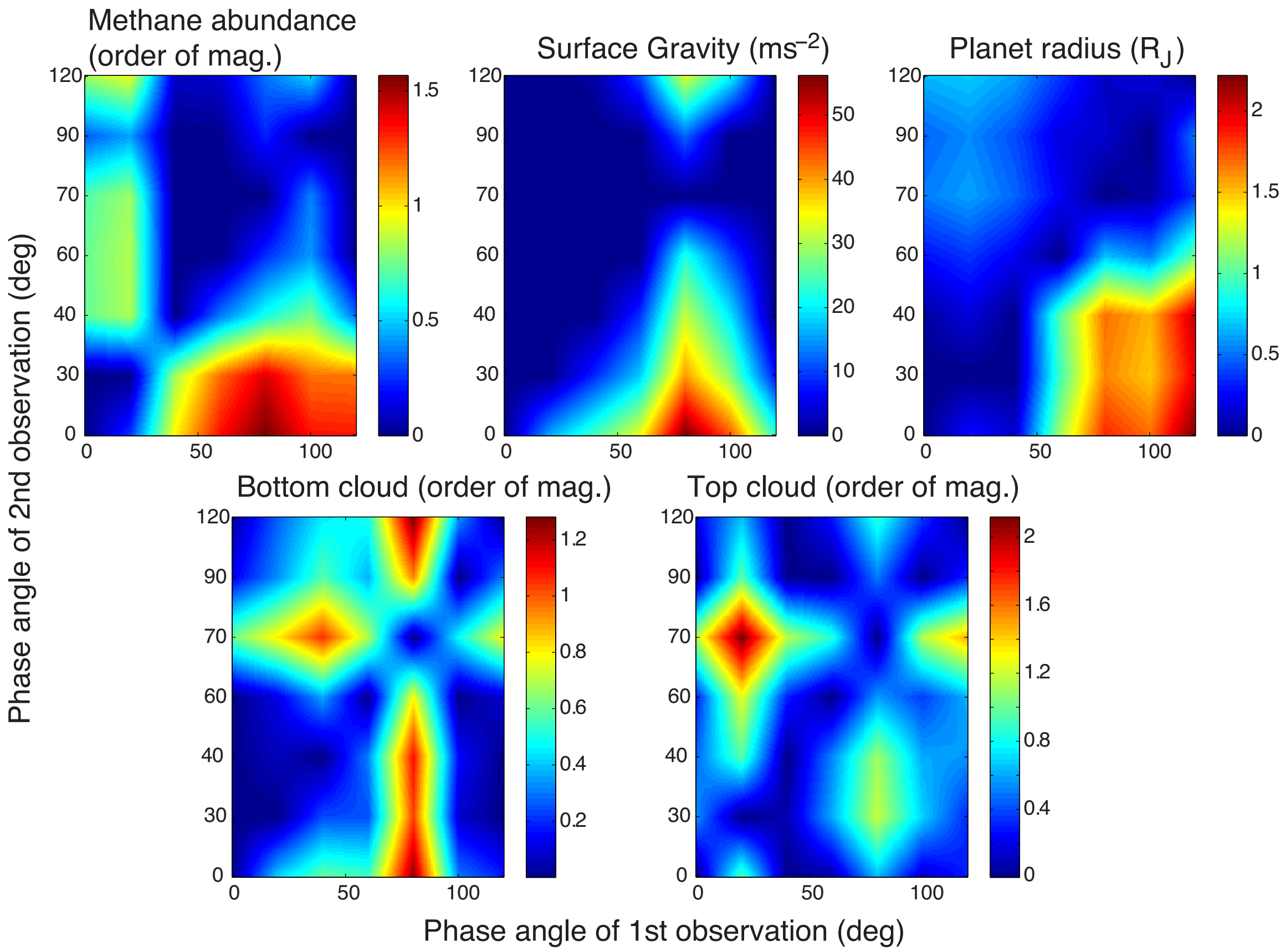

**Figure 20. Approximation of the improvement expected from two phase-varying observations over a single observation. Improvement in 68% confidence interval ranges for methane abundance, surface gravity, planet radius, retrieved phase angle and two-cloud top pressures are shown; warm colors represent increasing improvement between two observations taken at different phase angles. Results are for HD 192310c (10x metallicity case) at SNR = 5. The phase angles of the first and second observations are represented on the x- and y-axes respectively; units of the relative improvement (and colorbars) are indicated in the figure titles.**

Figure 20 shows that a steadily increasing improvement in estimates of planet radius can be expected with two observations at phase angles that exceed 45°. For example, for one observation at 70° and another at 120°, the estimate for planet radius can be improved by a factor of two, a significant improvement when considering that the best-fit solution from a single observation was already within a factor of two of the true solution. Similarly, confidence intervals on surface gravity may be improved by as much as 55 - 60 ms$^{-2}$ if observations are gathered at 0° and 70° phase angle, though gains of >30 ms$^{-2}$ are possible with other combinations. Both these cases were also illustrated in Figure 19.

Similarly, the uncertainty in methane composition from retrievals against an observation taken at near-quadrature can be driven down by almost two orders of magnitude if combined with a low phase angle observation, but in the absence of such an observation, may still be reduced by 0.8 orders with an observation at 45° phase angle. The intersection criterion presents a way to estimate trends in phase-varying behavior; future simultaneous retrievals and joint probability distributions created from multiple observations will quantify the exact improvement.

Such improvements will of course be reduced with increasing SNR (smaller probability bounds, greater overlap), and with more than two observations. During an actual space-based observational campaign, operational or other constraints may limit the ability to observe a planet at a favorable viewing geometry. Therefore, we now estimate the likelihood of improvement in confidence intervals.

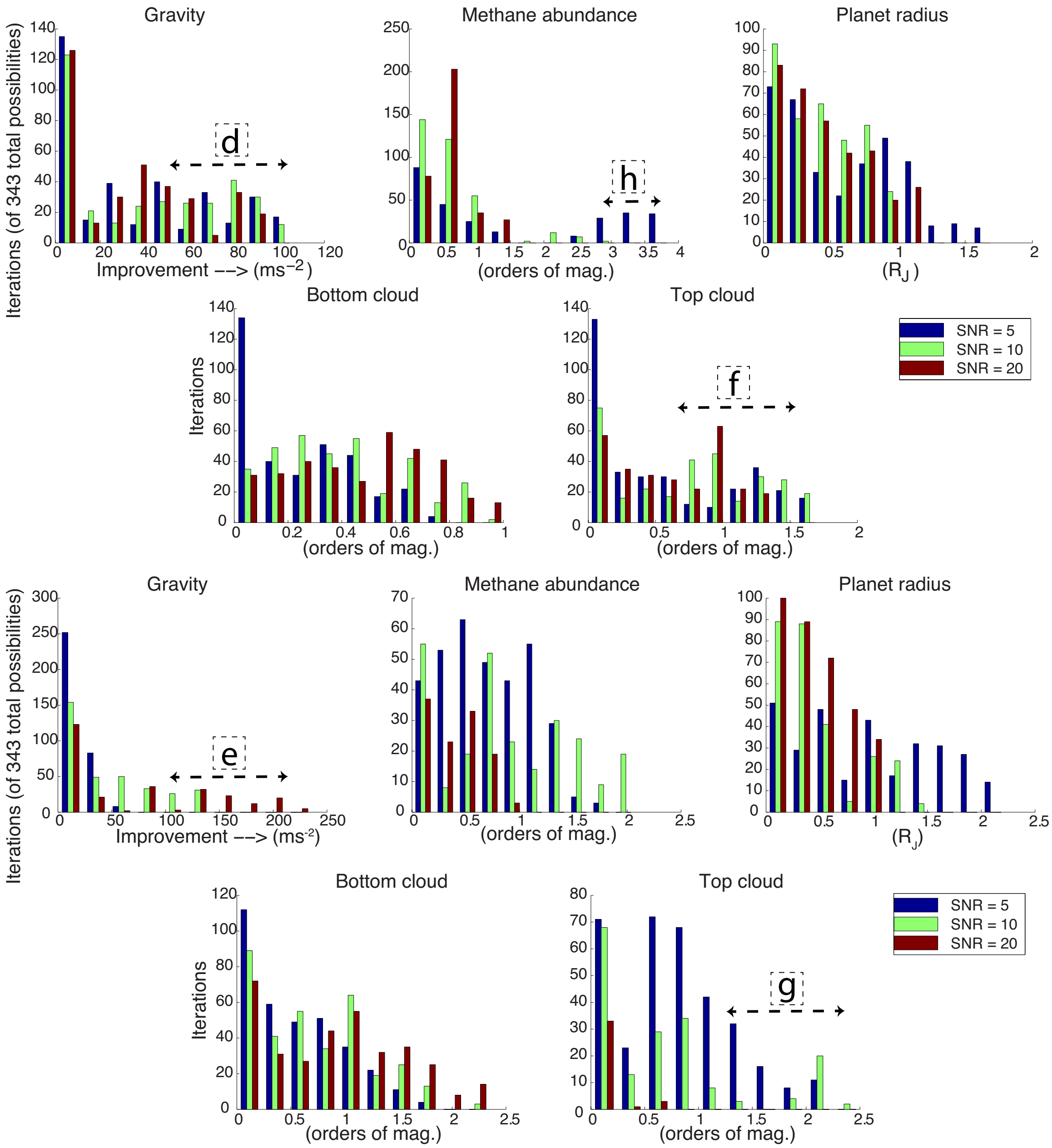

**Figure 21. Approximation of the improvement expected from three phase-varying observations over a single observation. Improvement in 68% confidence intervals for an intersection criterion applied to three randomly chosen phase-varying observations of (A) Planet A and (B) HD 192310c, 10x metallicity case, compared to a randomly chosen single observation. A total of 343 possible permutations (y-axis) are possible. SNR values of 5-20 are shown (blue: SNR 5, green: SNR 10, red: SNR 20). See text for references to text markers d-h. This plot may be used to determine trends for improvement in uncertainty estimates by retrieving against multiple datasets collected at differing phase angles. For example, for gravity, most cases show no improvement in gravity estimates (improvement clusters around 0 ms$^{-2}$), regardless of SNR, when confidence intervals from multiple observations are stacked together, regardless of the phase angle of the single observation. However, for methane abundance, a significant number of SNR 5 cases (blue) show at 0.5-1 order of magnitude improvement when the intersection criterion is applied to multiple observations. A similar trend is noted for higher SNR cases, although the number of cases that note this improvement drops off as expected.**

We randomly choose three phase-varying observations, and compare the intersection criterion result with that of a single observation, also randomly chosen.

Figure 21 plots the improvement with the intersection criterion, for every possible three-observation combination of the seven phase angles studied here (total of 343 possibilities), for both Planet A and the 10x HD 192310c case. As expected, for either planet, the improvement generally does not exceed an order of magnitude for the SNR = 20 case. However for SNR = 5, the improvement is significant in all cases, even for largely invariant parameters such as gravity (marker d-e), but particularly for cloud parameters (marker f-g) and methane abundance. Up to 3.5 orders of magnitude in improvement for the methane abundance is seen for Planet A (marker h), depending on the phase angle combinations; recall that for this case, we were unable to do much better than determine the presence of methane.

Similar results are seen for four and more observations. Such plots allow us to build an idea of trends for improvement in uncertainty estimates. These trends will be important for mission planning for WFIRST; these also present a starting point for estimating science return for realistic mission scenarios, where data is available from multiple observations at different phase angles and low SNR.

## 5. Conclusions

We have studied a number of retrievals on simulated phase-varying spectra, incorporating different metallicities, star-planet fluxes and signal-to-noise ratios. Specifically we presented results of how the unknown planet radius and potentially poorly known observer-planet-star phase angle can impact retrievals of atmospheric methane abundance, cloud properties and surface gravity, among others.

Given a varying planet phase, we find that the methane abundance can typically only be constrained to the correct order of magnitude at SNR of 20 or greater. For all three metallicities considered (1x, 10x, 50x solar), only the case with an SNR of 20 reproduces to the correct methane abundance at all phase angles. Low SNR cases merely identify the presence of methane, with the abundance being highly uncertain across several orders of magnitude, an important result for the design of future space-based missions such as WFIRST.

Surface gravity appears essentially unconstrained. The top cloud in a two-layer cloud model is well constrained, but is indeterminate on the pressure gap between cloud layers, indicating that a one-cloud model might be better suited to the examples in this paper. However our MCMC methods are able to return a solution for planet radius within a factor of two of the true value, even at low SNR values. Surprisingly, the confidence interval on the radius solution decreases with increasing phase angle. Since the atmosphere is highly forward scattering and molecular bands are weak at more crescent phases, clouds become less important. Retrievals for radius are consequently best at the highest phase angles.

We find that knowledge of the phase angle, and therefore its elimination as a free parameter, does not significantly improve estimates for methane abundance, cloud parameters or gravity. However it does improve the radius determination by a factor of two. On the other hand, with no prior knowledge of orbital parameters, the most we can confidently tell about the phase angle from retrieved results is whether it is high or low. Observations with comparable SNR and multiple phase angles can therefore be extremely valuable in determining both orbital characteristics, if unknown or uncertain, and narrowing down planet radius.

Finally, we find that simultaneous retrievals on observations taken at multiple phase angles holds promise for narrowing the solution space of best-fit models. We estimate this using an intersection criterion and find a steadily increasing improvement in estimates of all parameters, even for generally indeterminate parameters such as surface gravity and retrieved phase angle. For low SNR cases, estimates for methane abundance can be improved by as much as 1-2 orders of magnitude if multiple observations at different phase angles are gathered, a fact of interest when planning future space-based observational campaigns. However, it is important not to assign too much importance to this, since even though bounds on the solution may decrease, this does not guarantee the *accuracy* of the solution. At low SNRs, the recovered methane solution is far separated from the true value at all phase angles. The best that multiple observations at low SNR can hope to accomplish is the information content of one observation at high SNR.

Our group is continuing to pursue MCMC methods for application to reflected-light spectral data in the context of a wide range of future missions, including WFIRST. Future work will focus on a continued improved treatment of clouds and hazes, as well as Raman scattering, although we expect this latter effect to be minimal, since Raman scattering features are weak in the visible wavelengths (Karkoschka 1994). We will also pursue retrievals to determine the planetary temperature-pressure profile via the reflection spectrum. Finally, we will perform simultaneous retrievals on observations taken at multiple phase angles, an improvement on the intersection criterion approximation investigated here.

## Acknowledgements

The NASA High-End Computing (HEC) Program, through the NASA Advanced Supercomputing (NAS) division at the Ames Research Center, provided resources supporting this work. Results reported herein benefited from collaborations and/or information exchange within NASA's Nexus for Exoplanet System Science (NExSS) research coordination network sponsored by the NASA Science Mission Directorate. M.N. is supported by the National Defense Science and Engineering Graduate Fellowship (NDSEG, 32 CFR 168a) and Red Sky Research, LLC. MSM acknowledges the support of the WFIRST Preparatory Science Program. T.R. gratefully acknowledges support from NASA through the Sagan Fellowship executed by the NASA Exoplanet Science Institute. Opinions, interpretations and recommendations expressed are those of the authors and are not necessarily endorsed by the US Air Force or the Department of Defense.